\documentclass[journal]{IEEEtran}
\ifCLASSINFOpdf
\else
\fi

\usepackage[utf8]{inputenc}
\usepackage{amsmath,amssymb,amsfonts}
\usepackage{cite}

\usepackage{algorithmic}
\usepackage{stfloats}
\usepackage{graphicx}
\usepackage{textcomp}
\usepackage{xcolor}
\usepackage{color}
\usepackage{pifont}
\usepackage{amsthm}
\newtheorem{theorem}{Theorem}

\usepackage{hyperref}
\usepackage{mathrsfs}
\usepackage{booktabs}
\usepackage{hyperref}
\usepackage{cases}
\usepackage{comment}
\usepackage{empheq} 
\usepackage{caption}
\usepackage{subcaption}
\allowdisplaybreaks[4]

\usepackage{url}  
\usepackage{graphicx}  
\usepackage[ruled,vlined,linesnumbered]{algorithm2e}

\begin{document}
\captionsetup[figure]{labelfont={rm},labelformat={default},labelsep=period,name={Fig.}}
\let\sss= \scriptscriptstyle
\title{Robust Beamforming Design for Near-Field DMA-NOMA mmWave Communications With Imperfect Position Information}
%
%
%

\author{Yue Xiu$^*$,~\IEEEmembership{Member,~IEEE},~Yang Zhao$^*$,~\IEEEmembership{Member,~IEEE},~Songjie Yang,~\IEEEmembership{Student Member,~IEEE},\\
Yufeng Zhang,~\IEEEmembership{Member,~IEEE},~Dusit~Niyato,~\IEEEmembership{Fellow,~IEEE}, Hongyang Du,~\IEEEmembership{Member,~IEEE},\\Ning Wei,~\IEEEmembership{Member,~IEEE}\\
\thanks{
Y. Xiu, S. Yang, and N. Wei are with the National Key Laboratory of Science and Technology on Communications, University of Electronic Science and Technology of China, Chengdu 611731, China (e-mail: xiuyue12345678@163.com, yangsongjie@std.uestc.edu.cn, wn@uestc.edu.cn). 

Y. Zhao is with Nanyang Technological University, Singapore (e-mail:s180049@e.ntu.edu.sg).

Y. Zhang is with the Department of Electronic Engineering, Tsinghua University, Beijing 100084, China (e-mail: zhangyufeng@mail.tsinghua.edu.cn).

H. Du and D. Niyato are with the College of Computing and Data Science, Nanyang
Technological University, Singapore 639798 (e-mail: hongyang001@e.ntu.edu.sg, dniyato@ntu.edu.sg).
}
\thanks{The corresponding author is Ning Wei.}
\thanks{$*$ These authors contributed equally to this work.}}

\maketitle
\begin{abstract}
For millimeter-wave (mmWave) non-orthogonal multiple access (NOMA) communication systems, we propose an innovative near-field (NF) transmission framework based on dynamic metasurface antenna (DMA) technology. In this framework, a base station (BS) utilizes the DMA hybrid beamforming technology combined with the NOMA principle to maximize communication efficiency between near-field users (NUs) and far-field users (FUs). In conventional communication systems, obtaining channel state information (CSI) requires substantial pilot signals, significantly reducing system communication efficiency. We propose a beamforming design scheme based on position information to address with this challenge. This scheme does not depend on pilot signals but indirectly obtains CSI by analyzing the geometric relationship between user position information and channel models. However, in practical applications, the accuracy of position information is challenging to guarantee and may contain errors. We propose a robust beamforming design strategy based on the worst-case scenario to tackle this issue. Facing with the multi-variable coupled non-convex problems, we employ a dual-loop iterative joint optimization algorithm to update beamforming using block coordinate descent (BCD) and derive the optimal power allocation (PA) expression. We analyze its convergence and complexity to verify the proposed algorithm's performance and robustness thoroughly. We validate the theoretical derivation of the CSI error bound through simulation experiments. Numerical results show that our proposed scheme performs better than traditional beamforming schemes. Additionally, the transmission framework exhibits strong robustness to NU and FU position errors, laying a solid foundation for the practical application of mmWave NOMA communication systems. The NF transmission framework for mmWave NOMA communication systems based on DMA technology proposed in this work shows significant advantages in improving communication efficiency, reducing reliance on pilot signals, and coping with position errors. This provides new insights for the future development of mmWave communication technology.
\end{abstract}


%
\IEEEpeerreviewmaketitle

\section{Introduction}
\IEEEPARstart{I}{n} recent years, with the rapid development of communication technology, data traffic, and data terminals have increased rapidly, which leads to spectrum shortage\cite{6736750,9472958,7010531}. To address this problem, the emergence of millimeter wave (mmWave) and non-orthogonal multiple access (NOMA) technologies offers technical support for overcoming these challenges\cite{8294044,9472958}. Although mmWave and NOMA technologies have demonstrated significant advantages in improving spectrum efficiency, the high energy consumption and high hardware costs of mmWave-NOMA systems remain unresolved, restricting their widespread deployment\cite{9976948}. With the development of 6G communication, these problems are expected to become even more pronounced\cite{10054381}.

The rapid development of metamaterials offers technical support for addressing the issues above. Recently, researchers have proposed a new antenna model based on metamaterials, known as dynamic metasurface antennas (DMA)\cite{9141218,9324910}. This paper focuses on amplitude control DMA\cite{9141218}. The DMA generates reference electromagnetic waves through a feed source, propagating along a surface engraved with a specific beam pattern. This surface functions like a metasurface with an array of tiny antennas. As the electromagnetic waves propagate across this metasurface, they are eventually emitted from the radiating elements on the surface into free space. The DMA records the interference between the reference and desired waves during this process. This interference phenomenon is similar to the complex waveform generated when two sets of waves meet. By precisely controlling this interference, the DMA can adjust the intensity and phase of the reference wave. This precise control allows the DMA to generate reconfigurable beam patterns that can be dynamically adjusted, thereby enabling flexible spatial beamforming\cite{9762020,9392006}. Compared with traditional mmWave communication systems based on phase shift devices, DMA does not rely on phase shifters and amplifying circuits, resulting in lower cost and energy consumption. This provides a technical foundation for the widespread deployment of mmWave-NOMA systems. Studies have shown that to achieve optimal performance in DMA systems, it is essential to design and optimize both the phase and amplitude\cite{8756024,9991243,9272351}. This process requires obtaining sufficient instantaneous CSI from the BS as an initial step. Although various channel estimation techniques developed from pilot signals are commonly used to acquire instantaneous CSI\cite{10473671,10081022,2023arXiv230400440Y}, this process faces several challenges.
Firstly, traditional CSI estimation techniques consume significant pilot resources when there are many BS antenna arrays. Secondly, in real-world scenarios, CSI varies over time, and updating passive beamforming may lag due to training and communication delays, diminishing DMA's benefits. To tackle these challenges, a communication scheme based on user position information has recently been proposed\cite{10070578}. Specifically, in practical applications, the BS position is typically fixed and known after deployment, while user positions can be easily determined through various methods. Once the user position is identified, the channel matrix or vector can be directly reconstructed based on the spatial geometric relationship among the array responses of the BS, the user, and each terminal, thereby eliminating the need for traditional channel training.

However, localization systems usually do not accurately measure the user's position\cite{10438978}. Due to the limitations of positioning equipment or hardware issues, localization errors are inevitable, affecting the reconstructed channel by CSI uncertainty\cite{7442902}. This CSI uncertainty can seriously impact system performance, especially in demanding communication environments. Therefore, to ensure optimal system performance, the design of transmit or passive beamforming must be robust against CSI uncertainty. So far, some significant preliminary works have been on robust beamforming in mmWave communication systems. For example, in \cite{10552118}, the authors addressed CSI errors in mmWave systems by utilizing the singular value decomposition structure of the effective channel. They proposed a robust method for joint active and passive beamforming based on cascaded channel estimation, which showed clear advantages over compressed sensing-based methods. In \cite{10535255}, for mmWave systems, the authors enhanced the approach of using a deep learning model to reduce beam training overhead, ultimately achieving stability and robustness in dynamic environments. In \cite{raha2024advancing}, the authors considered the existence of CSI errors in cell-free mmWave multiple-input multiple-output (MIMO) networks. They proposed a robust algorithm based on Wiener filtering to address these errors. In \cite{huang2023robust}, for mmWave DMA systems, the authors considered both bounded channel interference errors and statistical channel interference errors, designing a robust transmission and passive beamforming optimization algorithm based on SCA and the penalized concave-convex procedure (CCCP). These studies provide valuable references for addressing CSI uncertainty. However, these studies did not consider CSI errors caused by user position uncertainty.
Moreover, these algorithms may experience slow convergence when the numbers of transmitting antennas or receiving elements increase. To our knowledge, the implementation of an effective position-based robust transmit and passive beamforming optimization method for mmWave DMA-NOMA systems has yet to be explored. This gap in the research motivates us to start this study.

This paper considers that DMA-enabled mmWave-NOMA systems need only position information and do not need to consume pilot resources to obtain CSI. Our research aims to address the CSI reconstruction error problem caused by user position uncertainty and improve the algorithm's convergence speed and system performance when many transmit antennas or receive elements are used. We develop a method to maximize the system's worst-case rate with user position error while guaranteeing fairness between NUs and FUs through robust beamforming design, enhancing the reliability and efficiency of mmWave DMA systems in practical applications. The main contributions of this work are summarized below.

\begin{itemize}
    \item First, we consider that the user positioning error is limited to the spherical region\cite{9416239}. According to the user positioning error bound, we derive the approximate CSI error bound. Then, we optimize the CSI error to to maximize the minimum system rate, that is, the worst-case rate. In this NOMA system, each NU and FU form a NOMA group, and the NU and FU within each group utilize a common hybrid beamformer. According to the designed beamformer, the problems of power allocation and robust beamforming are established to maximize the sum of the worst reachable rates while satisfying the quality of service requirements and SIC decoding constraints.
    \item After determining the CSI error bound, we formulate an optimization problem for robust transmission and passive beamforming under worst-case scenarios. Due to the non-convexity and complexity of solving the original problem, we propose a new iterative optimization method to achieve a near-optimal solution. Specifically, the internal minimization is performed using the BCD algorithm. Then, the outer maximization is carried out iteratively to find the optimal solution to the PA problem.
    \item Based on the proposed optimization method, we outline the complete algorithm and analyze its convergence performance and computational complexity. We then verify the theoretical derivation of the CSI error bound. Finally, numerical results are used to assess the robustness of the proposed algorithm. Compared to traditional non-robust methods, the proposed algorithm demonstrates greater resilience to CSI uncertainty.
\end{itemize}


\begin{table}[!ht]
\centering
\caption{Notations.}
\label{notations}
\begin{tabular}{cp{6.5cm}}
\toprule[1pt]
\textbf{Symbol} & \textbf{Descriptions} \\
\midrule
$\mathcal{H}(\mathcal{V})$ & The set of horizontal(vertical) DMA elements. \\
$\boldsymbol{p}_{N,i}(\boldsymbol{p}_{F,i})$ &  The real position vector of the $i$-th NU(FU). \\
$\hat{\boldsymbol{p}}_{N,i}(\hat{\boldsymbol{p}}_{F,i})$ & The  position estimation vector of the $i$-th NU(FU). \\
$\boldsymbol{\Delta}\boldsymbol{p}_{N,i}(\boldsymbol{\Delta}\boldsymbol{p}_{F,i})$ & The  position estimation error vector of the $i$-th NU(FU). \\
$\boldsymbol{h}_{N,i}(\boldsymbol{h}_{F,i})$ & The channel vector of the $i$-th NU(FU). \\
$\boldsymbol{h}_{\varepsilon,i}^{LoS}(\boldsymbol{h}_{\varepsilon,i}^{NLoS})$ & The LoS(NLoS) channel components of the $i$-th UE. \\
$\boldsymbol{\beta}_{N,i}(\boldsymbol{\beta}_{F,i})$ & The large-scale path loss coefficient vector of the $i$-th NU(FU). \\
$\boldsymbol{\alpha}_{N,i}(\boldsymbol{\alpha}_{F,i})$ & The array steering vector of the $i$-th NU(FU). \\
$\alpha$ & Path loss factor. \\
$\kappa_{R,N}(\kappa_{R,F})$ & NF(RF) Rician factor. \\
$P_{N,i}(P_{F,i})$ & PA factor for the $i$-th NU(FU). \\
$\boldsymbol{W}$ & Transmit beamforming. \\
$\boldsymbol{\upsilon}_{i}$ &  The $i$-th UE's baseband digital beamforming \\
$K$ & Number of NUs(FUs). \\
\bottomrule[1pt]
\end{tabular}\label{TA1}
\end{table}

\section{System Model and Problem Formulation}\label{II}

In this section, we introduce the mmWave DMA-NOMA system model. Specifically, in \textbf{Section~\ref{SA}}, we introduce the user position model and position error model. Then, based on the user position and position error models, we introduce the channel model and channel error in \textbf{Section~\ref{SB}}. Finally, based on the channel model, we introduce the received signal model and the mmWave DMA-NOMA worst-case robust optimization problem in \textbf{Section~\ref{SC}}. All the symbols of this paper are summarized in \textbf{Table~\ref{TA1}}.

\subsection{User position model and position error model}\label{SA}
\begin{figure}[htbp]
  \centering
  \includegraphics[scale=0.5]{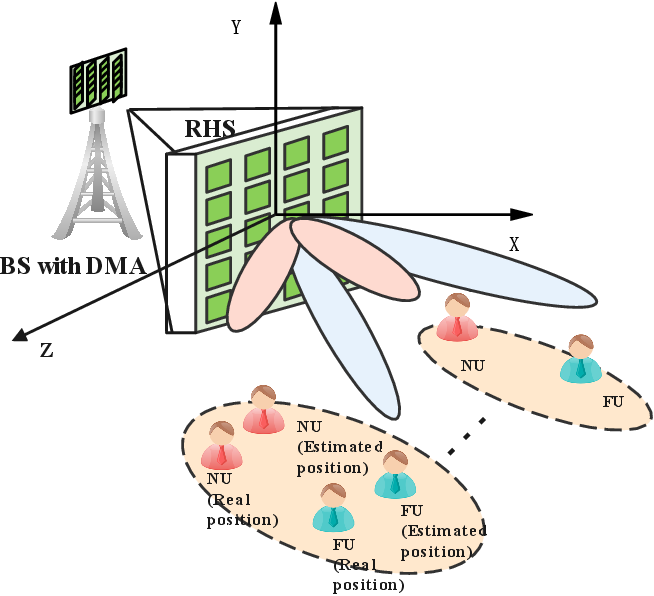}
  \caption{MmWave DMA-NOMA communication system in an
NF propagation environment. The NU and FU estimated positions are acquired from the existing localization systems are adopted to reconstruct the channel. }\vspace{-10pt}
\label{FIGURE0}
\end{figure}
This paper investigates the mmWave-NOMA communication system integrated within a three-dimensional (3D) near-field (NF) propagation environment. This system incorporates a DMA comprising $N_{T}=N_{T,v}N_{T,h}$ elements. As depicted in Fig.\ref{FIGURE0}, $N_{T,v}$ represents the number of vertical DMA elements, and $N_{T,h}$ represents the number of horizontal DMA elements. The sets $\mathcal{V}=\{1,\ldots, N_{T,v}\}$ and $\mathcal{H}=\{1,\ldots, N_{T,h}\}$ correspond to the vertical and horizontal DMA elements, respectively. In this system, BS simultaneously serves $K$ single-antenna near-field users (NUs) and $K$ single-antenna far-field users (FUs).  According to \cite{10068140}, the distance is less than the Rayleigh distance $2D^{2}/\bar{\lambda}$, called the NF, and the distance beyond the Rayleigh distance $2D^{2}/\bar{\lambda}$ is considered the far-field (FF). In this expression, $D$ represents the aperture of the BS, and $\bar{\lambda}$ denotes the wavelength of the mmWave.  Based on \cite{zhang2024near}, the origin of the 3D coordinate system is set as the central element of the DMA. The coordinate of the $v$-th row and the $h$-th column element of the DMA is denoted as $\boldsymbol{x}_{v,h}=\left(0,\lambda\left(\frac{2h-N_{T,v}-1}{4}\right),\lambda\left(\frac{2h-N_{T,h}-1}{4}\right)\right),v\in\mathcal{V}, h\in\mathcal{H}$.

The position coordinates of the $i$-th NU and FU are given by $\boldsymbol{p}_{N,i}=\left[p_{N,i,x},p_{N,i,y},p_{N,i,z}\right]^{\sss T}\in\mathbb{R}^{3\times 1}$ and $\boldsymbol{p}_{F,i}=\left[p_{F,i,x},p_{F,i,y},p_{F,i,z}\right]^{\sss T}\in\mathbb{R}^{3\times 1}$, respectively. The position estimation vectors of the $i$-th NU and FU are denoted as $\hat{\boldsymbol{p}}_{N,i}=\left[\hat{p}_{N,i,x},\hat{p}_{N,i,y},\hat{p}_{N,i,z}\right]^{\sss T}\in\mathbb{R}^{3\times 1}$ and $\hat{\boldsymbol{p}}_{F,i}=\left[\hat{p}_{F,i,x},\hat{p}_{F,i,y},\hat{p}_{F,i,z}\right]^{\sss T}\in\mathbb{R}^{3\times 1}$, respectively. The $i$-th NU's position error vector is $\boldsymbol{\Delta}\boldsymbol{p}_{N,i}=\left[\Delta p_{N,i,x},\Delta p_{N,i,y},\Delta p_{N,i,z}\right]^{\sss T}\in\mathbb{R}^{3\times 1}$, and the $i$-th FU's position error vector is $\boldsymbol{\Delta}\boldsymbol{p}_{F,i}=\left[\Delta p_{F,i,x},\Delta p_{F,i,y},\Delta p_{F,i,z}\right]^{\sss T}\in\mathbb{R}^{3\times 1}$.  According to \cite{6691955,5089957}, the user's position is obtained by using Ultra-wideband (UWB) or Global Positioning System (GPS), and the relationship among the position error vector, position estimation vector, and true position vector is $\boldsymbol{p}_{N,i}=\hat{\boldsymbol{p}}_{N,i}+\boldsymbol{\Delta}\boldsymbol{p}_{N,i}$ and $\boldsymbol{p}_{F,i}=\hat{\boldsymbol{p}}_{F,i}+\boldsymbol{\Delta}\boldsymbol{p}_{F,i}$, respectively.

\subsection{Channel model and channel error model}\label{SB}
According to\cite{zhang2024near}, the mmWave channel model between the BS with DMA and the $i$-th NU and the mmWave channel model between the BS with DMA and the $i$-th FU are modeled as
\begin{align}
&\boldsymbol{h}_{\varepsilon,i}=\sqrt{\frac{\kappa_{R,\varepsilon}}{(1+\kappa_{R,\varepsilon})}}\boldsymbol{h}_{\varepsilon,\iota}^{LoS}+\frac{1}{\sqrt{1+\kappa_{R,\varepsilon}}}\boldsymbol{h}_{\varepsilon,i}^{NLoS},\varepsilon\in\mathcal{C},\label{pro1}
\end{align}    
where $\mathcal{C}=\{N,F\}$ indicates the NF and FF, and $\kappa_{\sss R,\varepsilon}\in\mathbb{C}^{1\times1}$ represents the NF and FF Rician factors, respectively. The LoS components of the $i$-th NU and the $i$-th FU are $\boldsymbol{h}_{\varepsilon,i}^{\sss LoS}\in\mathbb{C}^{\sss N_{T}\times 1}$. The NLoS components of the $i$-th NU and the $i$-th FU are expressed as $\boldsymbol{h}_{\varepsilon,i}^{\sss NLoS}\in\mathbb{C}^{\sss N_{T}\times 1}$. Based on \cite{10078317}, $\boldsymbol{h}_{\varepsilon,i}^{\sss LoS}$ is denoted as
\begin{align}
\boldsymbol{h}_{\varepsilon,i}^{\sss LoS}=\boldsymbol{\beta}_{\varepsilon,i}\odot\boldsymbol{a}_{\varepsilon,i}, \varepsilon\in\mathcal{C},\label{pro2}
\end{align}
in which $\boldsymbol{\beta}_{\varepsilon,i}=\left[\sqrt{\beta_{1,1,i}^{\varepsilon}},\ldots,\sqrt{\beta_{N_{T,v},N_{T,h},i}^{\varepsilon}}\right]^{T}$ denotes large-scale path loss coefficient vectors.  The array steering vectors of the NF and FF channels $\boldsymbol{a}_{\varepsilon,i}$ are expressed as
\begin{align}
&\boldsymbol{a}_{\varepsilon,i}=\left[e^{-j\frac{2\pi}{\lambda}(\|\boldsymbol{x}_{1,1}-\boldsymbol{p}_{\varepsilon,i}\|)},\ldots,e^{-j\frac{2\pi}{\lambda}(\|\boldsymbol{x}_{N_{T,v},N_{T,h}}-\boldsymbol{p}_{\varepsilon,i}\|)}\right]^{\sss T}.\label{pro3}
\end{align}
The large-scale path loss coefficients from the DMA element situated at the $h$-th row and $v$-th column to the $i$-th NU and FU are given as follows:
\begin{align}
&\beta_{v,h,i}=\left(\frac{\zeta_{\varepsilon,0}}{d_{0}^{\alpha}}\right)\frac{1}{\left(d_{v+(h-1)N_{T,h}\rightarrow i}\right)^{\alpha}},~v\in\mathcal{V}, ~h\in\mathcal{H},\label{pro4}
\end{align}    
in which $d_{0}$ is the reference distance, and $\alpha$ denotes the path loss factor, and $d_{\varepsilon, v+(h-1)N_{T,h}\rightarrow i}=\|\boldsymbol{x}_{v,h}-\boldsymbol{p}_{\varepsilon,i}\|_{2}$. According to \cite{10070578}, the NLoS Rayleigh fading factor in the system is important, especially when the user is located in the FF. 
However, the NF NLoS Rayleigh fading factor can be ignored\cite{zhang2024near}.  Therefore, $\boldsymbol{h}_{i}^{\sss N}$ can be reformulated as
\begin{align}
&\boldsymbol{h}_{N,i}=\sqrt{\frac{\kappa_{R,N}}{(1+\kappa_{R,N})}}\boldsymbol{h}_{N,i}^{\sss LoS}.\label{pro5}
\end{align}    
Although $\boldsymbol{h}_{F,i}^{\sss NLoS}$ is random in the FF channel model, we can stabilize its $l_{2}$-norm within a short time slot using a pilot data sequence\cite{renyi2007probability}.  When the FU moves slowly, a small variation of the channel condition between the BS and the FU
allows the $l_{2}$-norm of 
$\boldsymbol{h}_{F,i}^{\sss NLoS}$ to be approximately regarded as a constant term within a short time slot\cite{10070578}. Therefore, we assume that $\|\boldsymbol{h}_{F,i}^{\sss NLoS}\|_{2}$  is known and $\|\boldsymbol{h}_{F,i}^{\sss NLoS}\|_{2}=\tau_{N,i}^{\sss NLoS}$. These preset values will assist in determining the CSI error range and optimizing beamforming. Consequently, when $\hat{\boldsymbol{p}}$ is utilized to reconstruct the LoS components from the DMA BS to the $i$-th FU channel, the overall CSI error between the actual and reconstructed channels can be expressed as follows:
\begin{align}
&\boldsymbol{\Delta}\boldsymbol{h}_{\varepsilon,i}=\boldsymbol{h}_{\varepsilon,i}-\hat{\boldsymbol{h}}_{\varepsilon,i}^{\sss LoS}=\begin{cases}
\boldsymbol{\Delta}\boldsymbol{h}_{N,i}^{\sss LoS}& \varepsilon=N\\
\boldsymbol{\Delta}\boldsymbol{h}_{F,i}^{\sss LoS}+\boldsymbol{\Delta}\boldsymbol{h}_{F,i}^{\sss NLoS}& \varepsilon=F
\end{cases},\label{pro6}
\end{align}    
in which $\boldsymbol{\Delta}\boldsymbol{h}_{\varepsilon,i}^{\sss LoS}=\sqrt{\frac{\kappa_{R,\varepsilon}}{(1+\kappa_{R,\varepsilon})}}\boldsymbol{h}_{\varepsilon,i}^{\sss LoS}-\hat{\boldsymbol{h}}_{\varepsilon,i}^{\sss LoS}$, $\boldsymbol{\Delta}\boldsymbol{h}_{F,i}^{\sss NLoS}=\frac{1}{\sqrt{1+\kappa_{R,F}}}\boldsymbol{h}_{F,i}^{\sss NLoS}$. $\hat{\boldsymbol{h}}_{\varepsilon,i}^{\sss LoS}\in\mathbb{C}^{N_{T}\times 1}$ represents the estimated LoS channel of the DMA BS connection to the channel of the $i$-th user located in the NF and FF, and they can be obtained by using $\hat{\boldsymbol{p}}_{\varepsilon,i}$. This can be determined as follows: 
\begin{align}
\hat{\boldsymbol{h}}_{\varepsilon,i}^{\sss LoS}=&\left(\sqrt{\hat{\beta}_{1,1,i}^{\varepsilon}}e^{j\frac{2\pi}{\lambda}\|\boldsymbol{x}_{1,1}-\hat{\boldsymbol{p}}_{\varepsilon,i}\|_{2}},\ldots,\sqrt{\hat{\beta}_{N_{T,v},N_{T,h},i}^{\varepsilon}}\right.\nonumber\\
&\left.e^{j\frac{2\pi}{\lambda}\|\boldsymbol{x}_{N_{T,v},N_{T,h}}-\hat{\boldsymbol{p}}_{\varepsilon,i}\|_{2}}\right)^{T},\label{pro7}
\end{align}
where $\hat{\beta}_{v,h,i}^{\varepsilon}$ is the reconstruction value based on the position information.
To enhance the robustness of the system, we consider optimizing $\boldsymbol{\Delta}\boldsymbol{h}_{\varepsilon,i}$ and $\boldsymbol{\Delta}\boldsymbol{p}_{\varepsilon,i}$ to achieve the worst sum rate of the system, followed by further optimizing $\{P_{N,i},P_{F,i}\}$,  $\boldsymbol{W}$ and $\{\boldsymbol{v}_{i}\}$ to maximize the worst sum rate.


\subsection{Signal model and problem formulation}\label{SC}
Based on\cite{10315058}, $2K$ user clusters are divided into seven NOMA pairs, with each pair consisting of an NU and an FU. Based on \cite{9472958}, we assume that the $i$-th NOMA group also consists of one NU and one FU. Let $\boldsymbol{v}_{i}$ denote the baseband digital beamforming vector assigned to the $i$-th NOMA group. $s_{\varepsilon,i}$ represents the signals allocated to the $i$-th NU and FU, respectively, while $\mathbb{E}\{|s_{\varepsilon,i}|^{2}\}=1$, $P_{N,i}$ and $P_{F,i}$ represent the transmission power allocated to the $i$-th NU and FU. The data signal from the BS is represented as
\begin{align}
&\boldsymbol{x}=\sum\nolimits_{i=1}^{K}\boldsymbol{W}\boldsymbol{v}_{i}\left(\sqrt{P_{N,i}}s_{N,i}+\sqrt{P_{F,i}}s_{F,i}\right). \label{pro8}
\end{align}
According to (\ref{pro8}), the signals transmitted from the BS and received by the $i$-th NU and FU can be expressed as follows:
\begin{align}
&\boldsymbol{y}_{N,i}=\underbrace{\boldsymbol{h}_{N,i}\boldsymbol{W}\boldsymbol{v}_{i}\sqrt{P_{N,i}}s_{N,i}}_{\text{NF~desired~signal}}+\underbrace{\boldsymbol{h}_{N,i}\boldsymbol{W}\boldsymbol{v}_{i}\sqrt{P_{F,i}}s_{F,i}}_{\text{NF~intra-group~interference}}+\nonumber\\
&\underbrace{\boldsymbol{h}_{N,i}\sum\nolimits_{j=1,j\neq i}^{K}\boldsymbol{W}\boldsymbol{v}_{j}\left(\sqrt{P_{N,i}}s_{N,i}+\sqrt{P_{F,i}}s_{F,i}\right)}_{\text{NF~inter-group~interference}}+n_{N,i},\nonumber\\
&\boldsymbol{y}_{F,i}=\underbrace{\boldsymbol{h}_{F,i}\boldsymbol{W}\boldsymbol{v}_{i}\sqrt{P_{N,i}}s_{F,i}}_{\text{FF~desired~signal}}+\underbrace{\boldsymbol{h}_{F,i}\boldsymbol{W}\boldsymbol{v}_{i}\sqrt{P_{F,i}}s_{N,i}}_{\text{FF~intra-group~interference}}+\nonumber\\
&\underbrace{\boldsymbol{h}_{F,i}\sum\nolimits_{j=1,j\neq i}^{K}\boldsymbol{W}\boldsymbol{v}_{j}\left(\sqrt{P_{F,i}}s_{F,i}+\sqrt{P_{N,i}}s_{N,i}\right)}_{\text{FF~inter-group~interference}}+n_{F,i},\label{pro9}
\end{align}
in which $n_{\varepsilon,i}\sim\mathcal{CN}(0,\sigma_{\varepsilon}^{2}), \varepsilon\in\mathcal{C}$ is AWGN. $\boldsymbol{W}^{[(v-1)h+h,k]}=e^{-j\boldsymbol{r}_{s}\boldsymbol{x}_{v,h}^{k}}(\mathrm{Re}(\Gamma_{i}(\boldsymbol{x}_{v,h}^{k},\boldsymbol{r}_{f}))+1)/2$ denotes the normalized radiation amplitude for the feed $k$ at the $v$-th row and the $h$-th column element. $\Gamma_{i}(\boldsymbol{x}_{v,h}^{k},\boldsymbol{r}_{f})=\Gamma_{r}(\boldsymbol{x}_{v,h}^{k},\boldsymbol{r}_{s})\Gamma_{o}(\boldsymbol{x}_{v,h},\boldsymbol{r}_{f})$ signifies the radiation pattern generated by a reference electromagnetic wave aligned with the direction of the target position\cite{zhang2024near}. In this context, $\Gamma_{r}(\boldsymbol{x}_{v,h}^{k},\boldsymbol{r}_{s})=e^{-j\boldsymbol{r}_{s}\boldsymbol{x}_{v,h}^{k}}$ and $\Gamma_{o}(\boldsymbol{x}_{v,h},\boldsymbol{r}_{f})=e^{-j\boldsymbol{r}_{f}\boldsymbol{x}_{v,h}}$ represent the reference waves for the $v$-th row and the $h$-th column elements and the reference waves for the same element, respectively. The SINR values for the $i$-th NU and FU are given in (\ref{pro10}) at the top of this page. According to the explanation in\cite{9047922}, considering the particle condition, the effective gain of FUs is much smaller than that of NUs, which ensures that SIC works effectively in the $i$-th NOMA group. Under these definitions, robust DMA-NOMA systems and rate optimization problems can be represented by 
\begin{figure*}[ht] 
\centering 
\vspace*{8pt} 
\begin{align}
&\mathrm{SINR}_{N,i}=\frac{P_{N,i}|(\hat{\boldsymbol{h}}_{N,i}+\boldsymbol{\Delta}\boldsymbol{h}_{N,i})^{H}\boldsymbol{W}\boldsymbol{v}_{i}|^{2}}{\sum_{j=1,j\neq i}^{K}\left(P_{N,j}|(\hat{\boldsymbol{h}}_{N,i}+\boldsymbol{\Delta}\boldsymbol{h}_{N,i})^{H}\boldsymbol{W}\boldsymbol{v}_{j}|^{2}+P_{F,j}|(\hat{\boldsymbol{h}}_{N,i}+\boldsymbol{\Delta}\boldsymbol{h}_{N,i})^{H}\boldsymbol{W}\boldsymbol{v}_{j}|^{2}\right)+\sigma_{N}^{2}},\nonumber\\
&\mathrm{SINR}_{F,i}=\frac{P_{F,i}|(\hat{\boldsymbol{h}}_{F,i}+\boldsymbol{\Delta}\boldsymbol{h}_{F,i})^{H}\boldsymbol{W}\boldsymbol{v}_{i}|^{2}}{P_{N,i}|(\hat{\boldsymbol{h}}_{F,i}+\boldsymbol{\Delta}\boldsymbol{h}_{F,i})^{H}\boldsymbol{W}\boldsymbol{v}_{i}|^{2}+\sum_{j=1,j\neq i}^{K}\left(P_{N,j}|(\hat{\boldsymbol{h}}_{F,i}+\boldsymbol{\Delta}\boldsymbol{h}_{F,i})^{H}\boldsymbol{W}\boldsymbol{v}_{j}|^{2}+P_{F,j}|(\hat{\boldsymbol{h}}_{F,i}+\boldsymbol{\Delta}\boldsymbol{h}_{F,i})^{H}\boldsymbol{W}\boldsymbol{v}_{j}|^{2}\right)+\sigma_{F}^{2}}.\label{pro10}
\end{align}
\hrulefill 
\end{figure*}
\begin{subequations}
\begin{align}
\max_{\boldsymbol{W},\boldsymbol{v}_{i},\atop{P_{N,i},P_{F,i}}}\min_{\boldsymbol{\Delta}\boldsymbol{h}_{\varepsilon,i}, \boldsymbol{\Delta}\boldsymbol{p}_{\varepsilon,i}}&~\sum\nolimits_{i=1}^{K}\omega\log_{2}\left(1+\mathrm{SINR}_{N,i}\right)+(1-\omega)\nonumber\\
&\log_{2}\left(1+\mathrm{SINR}_{F,i}\right),\label{pro11a}\\
\mbox{s.t.}~
&\|\boldsymbol{v}_{i}\|^{2}=1,~1\leq i\leq K,&\label{pro11b}\\
&\boldsymbol{W}_{v,h}\in [0,1],v\in\mathcal{V},h\in\mathcal{H},&\label{pro11c}\\
&\|\boldsymbol{\Delta}\boldsymbol{h}_{\varepsilon,i}\|_{2}\leq\epsilon_{\boldsymbol{\Delta}\boldsymbol{h}_{\varepsilon,i}}, \varepsilon\in\mathcal{C},&\label{pro11d}\\
&\|\boldsymbol{p}_{\varepsilon,i}\|_{2}\leq \epsilon_{\boldsymbol{\Delta}\boldsymbol{p}_{\varepsilon,i}}, \varepsilon\in\mathcal{C}, &\label{pro11e}\\
&|(\hat{\boldsymbol{h}}_{\varepsilon,i}+\boldsymbol{\Delta}\boldsymbol{h}_{\varepsilon,i})^{H}\boldsymbol{W}\boldsymbol{v}_{j}|^{2}\leq \epsilon_{\varepsilon,i,j}, i\neq j&\label{pro11f}\\
&\log_{2}(1+\mathrm{SINR}_{\varepsilon,i})\geq \gamma_{\varepsilon,i}, \varepsilon\in\mathcal{C},&\label{pro11g}\\
&\sum\nolimits_{i=1}^{K}P_{N,i}+P_{F,i}=P,&\label{pro11h}\\
&P_{N,i}>0, P_{F,i}>0,&\label{pro11k}
\end{align}\label{pro11}%
\end{subequations}    
where $\omega$ is the weight. Constraint (\ref{pro11b}) denotes the normalized power of the DMA transmit beamforming vector. (\ref{pro11c}) represents the amplitude control constraint at DMA. (\ref{pro11d}) ensures that the overall CSI error of NU and FU is within an uncertain range. (\ref{pro11e}) limits the position estimation error of the NU and FU to a deterministic region. (\ref{pro11f}) is the constraints of inter-group interference, and the boundaries are small enough; that is, $\epsilon_{\varepsilon, i,j}\rightarrow 0$. (\ref{pro11g}) is the quality of service (QoS) constraints for NU and FU. We note that since the overall CSI error boundaries $\epsilon_{\boldsymbol{\Delta}\boldsymbol{h}_{\varepsilon,i}}$, $\epsilon_{\varepsilon,i,j}$ in constraints (\ref{pro11d}) and (\ref{pro11f}) are still uncertain, we must first determine the problems of $\epsilon_{\boldsymbol{\Delta}\boldsymbol{h}_{\varepsilon,i}}$ and $\epsilon_{\varepsilon, i,j}$ before solving this problem (\ref{pro11}). In the next section, we elaborate on the process of deriving $\epsilon_{\boldsymbol{\Delta}\boldsymbol{h}_{\varepsilon,i}}$ and $\epsilon_{\varepsilon,i,j}$.  After deriving $\epsilon_{\boldsymbol{\Delta}\boldsymbol{h}_{\varepsilon,i}}$ and $\epsilon_{\varepsilon,i,j}$, we propose the BCD algorithm to solve problem (\ref{pro11}). Specifically, problem (\ref{pro11}) is divided into two subproblems, i.e., beamforming optimization subproblem and power allocation subproblem, they are given by
\begin{subequations}
\begin{align}
\max_{\boldsymbol{W},\boldsymbol{v}_{i}}\min_{\boldsymbol{\Delta}\boldsymbol{h}_{\varepsilon,i}, \boldsymbol{\Delta}\boldsymbol{p}_{\varepsilon,i}}&~\sum\nolimits_{i=1}^{K}\omega\log_{2}\left(1+\mathrm{SINR}_{N,i}\right)+(1-\omega)\nonumber\\
&\log_{2}\left(1+\mathrm{SINR}_{F, i}\right),\label{pro12a}\\
\mbox{s.t.}~
&(\ref{pro11b}),(\ref{pro11c}),(\ref{pro11d}),(\ref{pro11e}),(\ref{pro11f}),(\ref{pro11g}),&\label{pro12b}
\end{align}\label{pro12}%
\end{subequations}    
and
\begin{subequations}
\begin{align}
\max_{\atop{P_{N,i},P_{F,i}}}&~\sum\nolimits_{i=1}^{K}\omega\log_{2}\left(1+\mathrm{SINR}_{N,i}\right)+(1-\omega)\nonumber\\
&\log_{2}\left(1+\mathrm{SINR}_{F, i}\right),\label{pro13a}\\
\mbox{s.t.}~
&(\ref{pro11g}),(\ref{pro11h}),(\ref{pro11k}).&\label{pro13b}
\end{align}\label{pro13}%
\end{subequations} 
The proposed algorithms based on dual-loop iterations for problems (\ref{pro12}) and (\ref{pro13}) are introduced in detail in the next section.

\section{Proposed Algorithm for Beamforming Optimization Problem}\label{III}
We find that the problem in (\ref{pro12}) is non-convex. To deal with this problem, we introduce slacking variables $\Psi_{\varepsilon,i}$ and $\bar{\Psi}_{\varepsilon,i}$, problem (\ref{pro12}) can be rewritten as
\begin{subequations}
\begin{align}
\max_{{\boldsymbol{W},\boldsymbol{v}_{i},\atop{\bar{\Psi}_{\varepsilon,i},\Psi_{\varepsilon,i}}}}\min_{\boldsymbol{\Delta}\boldsymbol{h}_{\varepsilon,i}, \boldsymbol{\Delta}\boldsymbol{p}_{\varepsilon,i}}&~\sum_{i=1}^{K}\omega\log_{2}\left(1+\frac{\Psi_{N,i}}{\bar{\Psi}_{N,i}}\right)+(1-\omega)\nonumber\\
&\log_{2}\left(1+\frac{\Psi_{F,i}}{\bar{\Psi}_{F,i}}\right),\label{pro14a}\\
\mbox{s.t.}~
&(\ref{pro11b}),(\ref{pro11c}),(\ref{pro11d}),(\ref{pro11e}),(\ref{pro11f}),(\ref{pro11g}),&\label{pro14b}\\
&\Psi_{\varepsilon,i}\geq P_{\varepsilon,i}|(\hat{\boldsymbol{h}}_{\varepsilon,i}+\boldsymbol{\Delta}\boldsymbol{h}_{\varepsilon,i})^{H}\boldsymbol{W}\boldsymbol{v}_{i}|^{2},&\label{pro14c}\\
&\bar{\Psi}_{N,i}\leq\sum\nolimits_{j=1,j\neq i}^{K}\left(P_{N,j}|(\hat{\boldsymbol{h}}_{N,i}+\right.&\nonumber\\
&\left.\boldsymbol{\Delta}\boldsymbol{h}_{N,i})^{H}\boldsymbol{W}\boldsymbol{v}_{j}|^{2}+P_{F,j}|(\hat{\boldsymbol{h}}_{N,i}+\right.&\nonumber\\
&\left.\boldsymbol{\Delta}\boldsymbol{h}_{N,i})^{H}\boldsymbol{W}\boldsymbol{v}_{j}|^{2}\right)+\sigma^{2},\label{pro14e}\\
&\bar{\Psi}_{F,i}\leq\sum\nolimits_{j=1,j\neq i}^{K}\left(P_{N,j}|(\hat{\boldsymbol{h}}_{F,i}+\right.&\nonumber\\
&\left.\boldsymbol{\Delta}\boldsymbol{h}_{F,i})^{H}\boldsymbol{W}\boldsymbol{v}_{j}|^{2}+P_{F,j}|(\hat{\boldsymbol{h}}_{F,i}+\right.&\nonumber\\
&\left.\boldsymbol{\Delta}\boldsymbol{h}_{F,i})^{H}\right.\left.\boldsymbol{W}\boldsymbol{v}_{j}|^{2}\right)+\sigma^{2},&\label{pro14f}\\
&\frac{\Psi_{\varepsilon,i}}{\bar{\Psi}_{\varepsilon,i}}\geq 2^{\gamma_{\varepsilon,i}-1}.&\label{pro14g}
\end{align}\label{pro14}%
\end{subequations}    

\subsection{Derivation of \texorpdfstring{$\epsilon_{\boldsymbol{\Delta}\boldsymbol{h}_{\varepsilon,i}}$ and $\epsilon_{\varepsilon,i,j}$}{\textmu}}
This section focuses on deriving $\epsilon_{\boldsymbol{\Delta}\boldsymbol{h}_{\varepsilon,i}}$ and $\epsilon_{\varepsilon,i}$ in the constraints (\ref{pro11d}) and (\ref{pro11f}). Therefore, channel error $\|\boldsymbol{h}_{\varepsilon,i}^{\sss LoS}-\hat{\boldsymbol{h}}_{\varepsilon,i}^{\sss LoS}\|_{2}$ can be derived based on (\ref{pro11e}). Next, we begin to derivate $\|\boldsymbol{h}_{\varepsilon,i}^{\sss LoS}-\hat{\boldsymbol{h}}_{\varepsilon,i}^{\sss LoS}\|_{2}$, which can be written as
\begin{align}
&\|\boldsymbol{h}_{\varepsilon,i}^{\sss LoS}-\hat{\boldsymbol{h}}_{\varepsilon,i}^{\sss LoS}\|_{2}=\sqrt{\zeta_{0}d_{0}^{\alpha}\Xi(\Delta\boldsymbol{p}_{\varepsilon,i})},\label{pro15}
\end{align}   
where $\Xi(\Delta\boldsymbol{p}_{\varepsilon,i})$ is given in (\ref{pro16}) at the top of this page. $\Theta_{\varepsilon,v+(h-1)N_{T,h}}$ is denoted as 
\begin{figure*}[ht] 
\centering 
\vspace*{8pt} 
\begin{align}
\Xi(\Delta\boldsymbol{p}_{\varepsilon,i})&=\sum_{v=1}^{N_{T,v}}\sum_{h=1}^{N_{T,h}}
\left\{\left\|\boldsymbol{x}_{v+(h-1)N_{T,h}}-\hat{\boldsymbol{p}}_{\varepsilon,i}-\Delta\boldsymbol{p}_{\varepsilon,i}\right\|_{2}^{-\alpha}+\left\|\boldsymbol{x}_{v+(h-1)N_{T,h}}-\hat{\boldsymbol{p}}_{\varepsilon,i}\right\|_{2}^{-\alpha}\right\}\nonumber\\
&-2\sum_{v=1}^{N_{T,v}}\sum_{h=1}^{N_{T,h}}\left\{\left\|\boldsymbol{x}_{v+(h-1)N_{T,h}}-\hat{\boldsymbol{p}}_{\varepsilon,i}-\Delta\boldsymbol{p}_{\varepsilon,i}\right\|_{2}^{-\frac{\alpha}{2}}\right.\left.\left\|\boldsymbol{x}_{v+(h-1)N_{T,h}}-\hat{\boldsymbol{p}}_{\varepsilon,i}\right\|_{2}^{-\frac{\alpha}{2}}\cos\left(2\pi\Theta_{\varepsilon,v+(h-1)N_{T,h}}/\lambda\right)\right\}.\label{pro16}
\end{align}
\hrulefill 
\end{figure*}
\begin{align}
&\Theta_{\varepsilon,v+(h-1)N_{T,h}}=\left(\|\boldsymbol{x}_{v+(h-1)N_{T,h}}-\hat{\boldsymbol{p}}_{\varepsilon,i}-\boldsymbol{\Delta}\boldsymbol{p}_{\varepsilon,i}\|_{2}\right.\nonumber\\
&\left.\|\boldsymbol{x}_{v+(h-1)N_{T,h}}-\hat{\boldsymbol{p}}_{\varepsilon,i}\|_{2}\right)\approx\left(\frac{(\boldsymbol{x}_{v+(h-1)N_{T,h}}-\hat{\boldsymbol{p}}_{\varepsilon,i})^{T}}{\|\boldsymbol{x}_{v+(h-1)N_{T,h}}-\hat{\boldsymbol{p}}_{\varepsilon,i}\|_{2}}\right)\nonumber\\
&\boldsymbol{\Delta}\boldsymbol{p}_{\varepsilon,i}=\boldsymbol{\xi}_{v+(h-1)N_{T,h}}^{T}\boldsymbol{\Delta}\boldsymbol{p}_{\varepsilon,i}.\label{pro17}
\end{align}

Since our goal is to obtain the worst-case sum rate over $\boldsymbol{\Delta}\boldsymbol{h}_{\varepsilon,i}$ and $\boldsymbol{\Delta}\boldsymbol{p}_{\varepsilon,i}$, we further reduce the feasible set of constraints (\ref{pro11d}) and (\ref{pro11f}). According to the triangular inequality\cite{10070578}, the upper bound of $P_{\varepsilon,i}|(\hat{\boldsymbol{h}}_{\varepsilon,i}+\boldsymbol{\Delta}\boldsymbol{h}_{\varepsilon,i})^{\sss H}\boldsymbol{W}\boldsymbol{v}_{i}|^{2}$ is expressed as 
\begin{align}
&P_{\varepsilon,i}|\hat{\boldsymbol{h}}_{\varepsilon,i}\boldsymbol{W}\boldsymbol{v}_{i}|^{2}+P_{N,i}|(\boldsymbol{\Delta}\boldsymbol{h}_{\varepsilon,i})^{\sss H}\boldsymbol{W}\boldsymbol{v}_{i}|^{2}\geq P_{N,i}|(\boldsymbol{h}_{\varepsilon,i}\nonumber\\
&+\boldsymbol{\Delta}\boldsymbol{h}_{\varepsilon,i})^{\sss H}\boldsymbol{W}\boldsymbol{v}_{i}|^{2}. \label{pro18}
\end{align}   
We continue to use the Cauchy-Schwartz inequality\cite{10081022}, and the upper bound is further denoted as
\begin{align}
&P_{\varepsilon,i}|\hat{\boldsymbol{h}}_{\varepsilon,i}\boldsymbol{W}\boldsymbol{v}_{i}|^{2}+P_{\varepsilon,i}|(\boldsymbol{\Delta}\boldsymbol{h}_{\varepsilon,i})^{\sss H}\boldsymbol{W}\boldsymbol{v}_{i}|^{2}\leq P_{\varepsilon,i}|\hat{\boldsymbol{h}}_{\varepsilon,i}\boldsymbol{W}\boldsymbol{v}_{i}|^{2}+\nonumber\\
&P_{\varepsilon,i}|(\boldsymbol{\Delta}\boldsymbol{h}_{\varepsilon,i})^{\sss 
H}|^{2}|\boldsymbol{W}\boldsymbol{v}_{i}|^{2}=g(\boldsymbol{\Delta}\boldsymbol{h}_{\varepsilon,i}).\label{pro19}
\end{align}   
Constraint (\ref{pro14c}) is rewritten as
\begin{align}
\Psi_{\varepsilon,i}\geq g(\boldsymbol{\Delta}\boldsymbol{h}_{\varepsilon,i}). \label{ppro0}
\end{align}
Similarly, we deal with constraints (\ref{pro14e}) and (\ref{pro14f}), and use the triangular inequality in \cite{10070578} to obtain the lower bound of $(P_{N,j}|(\hat{\boldsymbol{h}}_{N,i}+\boldsymbol{\Delta}\boldsymbol{h}_{N,i})^{\sss H}\boldsymbol{W}\boldsymbol{v}_{j}|^{2}+P_{F,j}|(\hat{\boldsymbol{h}}_{N,i}+\boldsymbol{\Delta}\boldsymbol{h}_{N,i})^{\sss H}\boldsymbol{W}\boldsymbol{v}_{j}|^{2})+\sigma^{2}$ and $(P_{N,j}|(\hat{\boldsymbol{h}}_{F,i}+\boldsymbol{\Delta}\boldsymbol{h}_{F,i})^{\sss H}\boldsymbol{W}\boldsymbol{v}_{j}|^{2}+P_{F,j}|(\hat{\boldsymbol{h}}_{F,i}+\boldsymbol{\Delta}\boldsymbol{h}_{F,i})^{\sss H}\boldsymbol{W}\boldsymbol{v}_{j}|^{2})+\sigma^{2}$, and they are given by
\begin{align}
&h(|\boldsymbol{\Delta}\boldsymbol{h}_{N,i}|^{2})=P_{N,j}|\hat{\boldsymbol{h}}_{N,i}\boldsymbol{W}\boldsymbol{v}_{j}|^{2}-P_{N,j}|\boldsymbol{\Delta}\boldsymbol{h}_{N,i}|^{2}|\boldsymbol{W}\boldsymbol{v}_{j}|^{2}+\nonumber\\
&P_{F,j}|\hat{\boldsymbol{h}}_{N,i}\boldsymbol{W}\boldsymbol{v}_{j}|^{2}-P_{F,j}|\boldsymbol{\Delta}\boldsymbol{h}_{N,i}|^{2}|\boldsymbol{W}\boldsymbol{v}_{j}|^{2},\label{pro21}\\
&h(|\boldsymbol{\Delta}\boldsymbol{h}_{F,i}|^{2})=P_{N,j}|\hat{\boldsymbol{h}}_{F,i}\boldsymbol{W}\boldsymbol{v}_{j}|^{2}-P_{N,j}|\boldsymbol{\Delta}\boldsymbol{h}_{F,i}|^{2}|\boldsymbol{W}\boldsymbol{v}_{j}|^{2}+\nonumber\\
&P_{F,j}|\hat{\boldsymbol{h}}_{F,i}\boldsymbol{W}\boldsymbol{v}_{j}|^{2}-P_{F,j}|\boldsymbol{\Delta}\boldsymbol{h}_{F,i}|^{2}|\boldsymbol{W}\boldsymbol{v}_{j}|^{2}.\label{pro22}
\end{align}   
To enhance the robustness of the problem (\ref{pro11}), we utilize (\ref{pro14e}) and (\ref{pro14f}) to further refine the feasible set of the constraints. Thus, (\ref{pro14e}) and (\ref{pro14f}) can be rewritten as follows
\begin{align}
&\bar{\Psi}_{\varepsilon,i}\leq\sum\nolimits_{j=1,j\neq i}^{K}h(|\boldsymbol{\Delta}\boldsymbol{h}_{\varepsilon,i}|^{2})+\sigma^{2}.&\label{ppro1}
\end{align}
Since $\epsilon_{\varepsilon,i,j}$ and $\epsilon_{\boldsymbol{\Delta}\boldsymbol{h}_{\varepsilon,i}}$ are the interference constraints and channel estimation error constraints, if the theoretically derived $\epsilon_{\varepsilon,i,j}$ and $\epsilon_{\boldsymbol{\Delta}\boldsymbol{h}_{\varepsilon,i}}$ are larger than the practical $\epsilon_{\varepsilon,i,j}$ and $\epsilon_{\boldsymbol{\Delta}\boldsymbol{h}_{\varepsilon,i}}$, the worst-case CSI experienced during the optimization process will become even worse than the practical worst-case CSI, hence resulting in a more robust solution for problem (\ref{pro11}). 
A smaller feasible set implies that there are fewer possible solutions to consider. This reduces the sensitivity of the problem to variations or uncertainties in the input data. With fewer options, the impact of data fluctuations on the optimal solution is minimized, leading to more consistent performance\cite{10438978}. Thus, to preserve the transmit and passive beamforming robustness, we let $\epsilon_{\varepsilon,i,j}=\max_{\boldsymbol{\Delta}\boldsymbol{p}_{\varepsilon,i}}g(\boldsymbol{\Delta}\boldsymbol{h}_{\varepsilon,i})$, $\epsilon_{\boldsymbol{\Delta}\boldsymbol{h}_{\varepsilon,i}}=\max_{\boldsymbol{\Delta}\boldsymbol{p}_{\varepsilon,i}}\|\boldsymbol{\Delta}\boldsymbol{h}_{\varepsilon,i}\|_{2}$. In addition, since $g(\boldsymbol{\Delta}\boldsymbol{h}_{\varepsilon,i})$ is a linear and monotonous increase over $|(\boldsymbol{\Delta}\boldsymbol{h}_{\varepsilon,i})^{\sss 
H}|^{2}$ when we maximize $|(\boldsymbol{\Delta}\boldsymbol{h}_{\varepsilon,i})^{\sss 
H}|^{2}$, this means that the feasible set of constraint (\ref{ppro0}) will be further narrowed, which will make problem (\ref{pro11}) more robust. Similarly, since $h(\boldsymbol{\Delta}\boldsymbol{h}_{\varepsilon,i})$ is a linear and monotonous decrease over $|(\boldsymbol{\Delta}\boldsymbol{h}_{\varepsilon,i})^{\sss 
H}|^{2}$ when we maximize $|(\boldsymbol{\Delta}\boldsymbol{h}_{\varepsilon,i})^{\sss 
H}|^{2}$, this means that the feasible set of constraints (\ref{ppro1}) will be further narrowed, which will also make problem (\ref{pro11}) more robust. Finally, we only need to maximize CSI error $|(\boldsymbol{\Delta}\boldsymbol{h}_{\varepsilon,i})^{\sss 
H}|^{2}$, and we find that $|(\boldsymbol{\Delta}\boldsymbol{h}_{\varepsilon,i})^{\sss 
H}|^{2}$ is over $\Xi(\Delta\boldsymbol{p}_{\varepsilon,i})$, the maximization problem of $|(\boldsymbol{\Delta}\boldsymbol{h}_{\varepsilon,i})^{\sss 
H}|^{2}$ is equivalent to maximize $\Xi(\Delta\boldsymbol{p}_{\varepsilon,i})$. Thus, we need to resolve the following optimization problems
\begin{subequations}
\begin{align}
\max_{\boldsymbol{\Delta}\boldsymbol{p}_{N,i},\boldsymbol{\Delta}\boldsymbol{p}_{F,i}}&~\sum\nolimits_{i=1}^{K}\Xi(\boldsymbol{\Delta}\boldsymbol{p}_{N,i})+\Xi(\boldsymbol{\Delta}\boldsymbol{p}_{F,i}),\label{pro23a}\\
\mbox{s.t.}~
&\|\boldsymbol{\Delta}\boldsymbol{p}_{\varepsilon,i}\|_{2}\leq \epsilon_{\boldsymbol{\Delta}\boldsymbol{p}_{\varepsilon,i}}. &\label{pro23b}
\end{align}\label{pro23}%
\end{subequations}%
\begin{theorem}\label{th1}
Problem (\ref{pro23}) is equivalently expressed as
\begin{subequations}
\begin{align}
\max_{\boldsymbol{\Delta}\boldsymbol{p}_{\varepsilon,i},\varpi_{\varepsilon,i},\hat{\varpi}_{\varepsilon,i}}&~\sum_{i=1}^{K}(4\pi^{4}/3\lambda^{4}N)((\varpi_{N,i})^{2}+(\varpi_{F,i})^{2})\nonumber\\
&+\hat{\varpi}_{N,i}+\hat{\varpi}_{F,i},\label{pro24a}\\
\mbox{s.t.}~
&(\ref{pro23b}),&\\
&\varpi_{\varepsilon,i}\geq(\boldsymbol{\Delta}\boldsymbol{p}_{\varepsilon,i})^{H}\boldsymbol{\Omega}_{\varepsilon,i}\boldsymbol{\Delta}\boldsymbol{p}_{\varepsilon,i},
&\label{pro24b}\\
&\hat{\varpi}_{\varepsilon,i}\geq(\boldsymbol{\Delta}\boldsymbol{p}_{\varepsilon,i})^{H}\boldsymbol{\Upsilon}_{\varepsilon,i}\boldsymbol{\Delta}\boldsymbol{p}_{\varepsilon,i},
&\label{pro24c}\\
&\varpi_{\varepsilon,i}\leq2\mathrm{Re}\{(\boldsymbol{\Delta}\boldsymbol{p}_{\varepsilon,i,0})^{H}\boldsymbol{\Omega}_{\varepsilon,i}\boldsymbol{\Delta}\boldsymbol{p}_{\varepsilon,i,0}\}&\nonumber\\
&-(\boldsymbol{\Delta}\boldsymbol{p}_{\varepsilon,i,0})^{H}\boldsymbol{\Omega}_{\varepsilon,i}\boldsymbol{\Delta}\boldsymbol{p}_{\varepsilon,i},&\\
&\hat{\varpi}_{\varepsilon,i}\leq2\mathrm{Re}\{(\boldsymbol{\Delta}\boldsymbol{p}_{\varepsilon,i,0})^{H}\boldsymbol{\Upsilon}_{\varepsilon,i}\boldsymbol{\Delta}\boldsymbol{p}_{\varepsilon,i,0}\}&\nonumber\\
&-(\boldsymbol{\Delta}\boldsymbol{p}_{\varepsilon,i,0})^{H}\boldsymbol{\Upsilon}_{\varepsilon,i}\boldsymbol{\Delta}\boldsymbol{p}_{\varepsilon,i},&
\end{align}\label{pro24}%
\end{subequations}%
where the expressions of $\boldsymbol{\Omega}_{\varepsilon,i}$ and $\boldsymbol{\Upsilon}_{\varepsilon,i}$ are given in (\ref{pro25}) at the bottom of the next page. This proof is given in \textbf{Appendix}~\textbf{A}. 
\end{theorem}
\begin{figure*}[hb] 
\hrulefill 
\centering 
\vspace*{8pt} 
\begin{align}
&\boldsymbol{\Omega}_{\varepsilon,i}=\sum\nolimits_{v=1}^{N_{T,v}}\sum\nolimits_{h=1}^{N_{T,h}}\left\{\left(\|\boldsymbol{v}_{v+(h-1)N_{T,h}}-\hat{\boldsymbol{p}}_{\varepsilon,i}\|_{2}-\epsilon_{\boldsymbol{\Delta}\boldsymbol{p}_{\varepsilon,i}}\right)^{-\frac{\alpha}{4}}\|\boldsymbol{v}_{v+(h-1)N_{T,h}}-\hat{\boldsymbol{p}}_{\varepsilon,i}\|_{2}^{-\frac{\alpha}{4}}\boldsymbol{\Xi}_{v+(h-1)N_{T,h}}\right\},\nonumber\\
&\boldsymbol{\Upsilon}_{\varepsilon,i}=\sum\nolimits_{v=1}^{N_{T,v}}\sum\nolimits_{h=1}^{N_{T,h}}\left\{\frac{1}{2}\boldsymbol{G}_{\varepsilon, \alpha}-\|\boldsymbol{v}_{v+(h-1)N_{T,h}}-\hat{\boldsymbol{p}}_{\varepsilon,i}\|_{2}^{-\frac{\alpha}{2}}\boldsymbol{G}_{\varepsilon,\frac{\alpha}{2}}+\frac{4\pi^{2}}{\lambda^{2}}\left(\|\boldsymbol{v}_{v+(h-1)N_{T,h}}-\hat{\boldsymbol{p}}_{\varepsilon,i}\|_{2}-\epsilon_{\boldsymbol{\Delta}\boldsymbol{p}_{\varepsilon,i}}\right)^{-\frac{\alpha}{4}}\right.\nonumber\\
&~~~~~~~~~\left.\|\boldsymbol{v}_{v+(h-1)N_{T,h}}-\hat{\boldsymbol{p}}_{\varepsilon,i}\|_{2}^{-\frac{\alpha}{2}}\boldsymbol{\Xi}_{N,v+(h-1)N_{T,h}}\right\},\nonumber\\
&\boldsymbol{G}_{\varepsilon, \alpha}=\alpha(\alpha+2)\|\boldsymbol{v}_{v+(h-1)N_{T,h}}-\hat{\boldsymbol{p}}_{\varepsilon,i}\|_{2}^{-\alpha-4}(\hat{\boldsymbol{p}}_{\varepsilon,i}-\boldsymbol{v}_{v+(h-1)N_{T,h}})(\hat{\boldsymbol{p}}_{\varepsilon,i}-\boldsymbol{v}_{v+(h-1)N_{T,h}})^{T}-\alpha\|\boldsymbol{v}_{v+(h-1)N_{T,h}}-\hat{\boldsymbol{p}}_{\varepsilon,i}\|_{2}^{-\alpha-2}\boldsymbol{I},\nonumber\\
&\boldsymbol{G}_{\varepsilon,\frac{\alpha}{2}}=\frac{\alpha}{2}(\frac{\alpha}{2}+2)\|\boldsymbol{v}_{v+(h-1)N_{T,h}}-\hat{\boldsymbol{p}}_{\varepsilon,i}\|_{2}^{-\frac{\alpha}{2}-4}(\hat{\boldsymbol{p}}_{\varepsilon,i}-\boldsymbol{v}_{v+(h-1)N_{T,h}})(\hat{\boldsymbol{p}}_{\varepsilon,i}-\boldsymbol{v}_{v+(h-1)N_{T,h}})^{T}-\frac{\alpha}{2}\|\boldsymbol{v}_{v+(h-1)N_{T,h}}-\hat{\boldsymbol{p}}_{\varepsilon,i}\|_{2}^{-\frac{\alpha}{2}-2}\boldsymbol{I},\nonumber\\
&\boldsymbol{\Xi}_{v+(h-1)N_{T,h}}=\boldsymbol{\eta}_{v+(h-1)N_{T,h}}\boldsymbol{\eta}^{T}_{v+(h-1)N_{T,h}},\label{pro25}
\end{align}
\hrulefill 
\end{figure*}


\subsection{Determination of \texorpdfstring{$\Delta\boldsymbol{h}_{N,i}$ and $\Delta\boldsymbol{h}_{F,i}$}{\textmu}}
After obtained the maximum value of problem (\ref{pro23}), based on (\ref{pro15}) we can calculate $\|\boldsymbol{h}_{\varepsilon,i}^{\sss LoS}-\hat{\boldsymbol{h}}_{\varepsilon,i}^{\sss LoS}\|_{2}$. Based on (\ref{pro26}), we consequently can determine the values of $\hat{f}(\|\boldsymbol{\Delta}\boldsymbol{h}_{N,i}^{\sss LoS}\|_{2})$ and $\hat{g}(\|\boldsymbol{\Delta}\boldsymbol{h}_{F,i}^{\sss LoS}\|_{2})$. Let $\hat{f}(\|\boldsymbol{\Delta}\boldsymbol{h}_{N,i}^{\sss LoS}\|_{2})=\epsilon_{\boldsymbol{\Delta}\boldsymbol{h}_{N,i}}$ and $\hat{g}(\|\boldsymbol{\Delta}\boldsymbol{h}_{F,i}^{\sss LoS}\|_{2})=\epsilon_{\boldsymbol{\Delta}\boldsymbol{h}_{F,i}}$. According to $g(\boldsymbol{\Delta}\boldsymbol{h}_{F,i})$ and $g(\boldsymbol{\Delta}\boldsymbol{h}_{F,i})$, we can determine $\epsilon_{\varepsilon,i,j}$. Since $\epsilon_{\varepsilon,i,j}$ has been obtained, we are now ready to determine $\boldsymbol{\Delta}\boldsymbol{h}_{\varepsilon,i}$. The optimization problem is rewritten as
\begin{figure*}[htbp] 
\centering 
\vspace*{8pt} 
\begin{align}
&\|\boldsymbol{\Delta}\boldsymbol{h}_{N,i}\|_{2}=\|\boldsymbol{\Delta}\boldsymbol{h}_{N,i}^{\sss LoS}\|_{2}=\left\|\sqrt{\frac{\kappa_{R,N}}{(1+\kappa_{R,N})}}\left(\boldsymbol{h}_{N,i}^{\sss LoS}-\hat{\boldsymbol{h}}_{N,i}^{\sss LoS}\right)\right\|_{2}\leq 
\sqrt{\frac{\kappa_{R,N}}{(1+\kappa_{R,N})}}\left\|\boldsymbol{h}_{N,i}^{\sss LoS}-\hat{\boldsymbol{h}}_{N,i}^{\sss LoS}\right\|_{2}\leq \hat{f}(\|\boldsymbol{\Delta}\boldsymbol{h}_{N,i}^{\sss LoS}\|_{2}),\nonumber\\
&\|\boldsymbol{\Delta}\boldsymbol{h}_{F,i}\|_{2}\leq\|\boldsymbol{\Delta}\boldsymbol{h}_{i}^{\sss LoS}\|_{2}+\|\boldsymbol{\Delta}\boldsymbol{h}_{i}^{\sss NLoS}\|_{2}=\left\|\sqrt{\frac{\kappa_{R,F}}{(1+\kappa_{R,F})}}(\boldsymbol{h}_{i}^{\sss LoS}-\hat{\boldsymbol{h}}_{i}^{\sss LoS})+\left(\sqrt{\frac{\kappa_{R,F}}{(1+\kappa_{R,F})}}-1\right)\hat{\boldsymbol{h}}_{i}^{\sss LoS}\right\|_{2}\nonumber\\
&+\frac{\delta_{i}^{\sss NLoS}}{\sqrt{1+\kappa_{R,F}}}\leq 
\sqrt{\frac{\kappa_{R,F}}{(1+\kappa_{R,F})}}\|\boldsymbol{h}_{i}^{\sss LoS}-\hat{\boldsymbol{h}}_{i}^{\sss LoS}\|_{2}+\left(1- 
\sqrt{\frac{\kappa_{R,F}}{(1+\kappa_{R,F})}}\right)\|\hat{\boldsymbol{h}}_{i}^{\sss LoS}\|_{2}+\frac{\delta_{i}^{\sss NLoS}}{\sqrt{1+\kappa_{R}}}\leq \hat{g}(\|\boldsymbol{\Delta}\boldsymbol{h}_{F,i}^{LoS}\|_{2}).\label{pro26}
\end{align}
\hrulefill 
\end{figure*}
\begin{subequations}
\begin{align}
\min_{\boldsymbol{\Delta}\boldsymbol{h}_{\varepsilon,i}}&~\sum_{i=1}^{K}\omega\log_{2}\left(1+\frac{\Psi_{N,i}}{\bar{\Psi}_{N,i}}\right)+(1-\omega)\nonumber\\
&\log_{2}\left(1+\frac{\Psi_{F,i}}{\bar{\Psi}_{F,i}}\right),\label{pro27a}\\
\mbox{s.t.}~
&(\ref{pro11d}),(\ref{pro11f}),(\ref{pro11g}),(\ref{pro14c}),(\ref{pro14e}),(\ref{pro14f}),&\label{pro27b}
\end{align}\label{pro27}%
\end{subequations}    
where the constraints (\ref{pro14e}) and (\ref{pro14f}) are non-convex, we use the SCA method to deal with (\ref{pro14e}) and (\ref{pro14f}), and the upper bound of the right-hand side of (\ref{pro14e}) and (\ref{pro14f}) is given by using the first-order Taylor series approximation, and (\ref{pro14e}) and (\ref{pro14f}) are transformed as
\begin{align}
&\bar{\Psi}_{\varepsilon,i}\leq\sum\nolimits_{j=1,j\neq i}^{K}2P_{\varepsilon,j}\mathrm{Re}\left\{\left(\left(\hat{\boldsymbol{h}}_{\varepsilon,i}+(\boldsymbol{\Delta}\boldsymbol{h}_{\varepsilon,i})^{(t-1)}\right)^{H}\boldsymbol{W}\boldsymbol{v}_{j}\right)\right.\nonumber\\
&\left.\boldsymbol{v}_{j}^{H}\boldsymbol{W}^{H}\boldsymbol{\Delta}\boldsymbol{h}_{\varepsilon,i}\right\}-\left|\left(\hat{\boldsymbol{h}}_{\varepsilon,i}+(\boldsymbol{\Delta}\boldsymbol{h}_{\varepsilon,i})^{(t-1)}\right)^{H}\boldsymbol{W}\boldsymbol{v}_{j}\right|^{2}.&\label{pro28}%
\end{align}
Substituting (\ref{pro28}) into problem (\ref{pro27}), problem (\ref{pro27}) is convex and we use CVX tool to solve problem (\ref{pro27}).

\subsection{Transmitted beamforming optimization}
With the fixed $P_{\varepsilon,i}$, $\boldsymbol{W}$, $\boldsymbol{\Delta}\boldsymbol{h}_{\varepsilon,i}$, and define $\boldsymbol{V}_{i}=\boldsymbol{v}_{i}\boldsymbol{v}_{i}^{H}$ and $\mathrm{rank}(\boldsymbol{V}_{i})=1$. We use the SDR method to relax constraint $\mathrm{rank}(\boldsymbol{V}_{i})=1$, problem (\ref{pro14}) is rewritten as
\begin{subequations}
\begin{align}
\max_{\{\boldsymbol{V}_{i}\},\Psi_{\varepsilon,i},\bar{\Psi}_{\varepsilon,i}}&~\sum_{i=1}^{K}\omega\log_{2}\left(1+\frac{\Psi_{N,i}}{\bar{\Psi}_{N,i}}\right)+(1-\omega)\nonumber\\
&\log_{2}\left(1+\frac{\Psi_{F,i}}{\bar{\Psi}_{F,i}}\right),\label{pro29a}\\
\mbox{s.t.}~
&\mathrm{tr}(\boldsymbol{V}_{i})=1,&\label{pro29b}\\
&\mathrm{tr}(\boldsymbol{\Phi}_{\varepsilon,i}\boldsymbol{V}_{j})\leq \epsilon_{\varepsilon,i,j},&\label{pro29c}\\
&\Psi_{\varepsilon,i}\geq P_{\varepsilon,i}\mathrm{tr}(\boldsymbol{\Phi}_{\varepsilon,i}\boldsymbol{V}_{i}),&\label{pro29d}\\
&\bar{\Psi}_{N,i}\leq\sum\nolimits_{j=1,j\neq i}^{K}\mathrm{tr}((P_{N,j}\boldsymbol{\Phi}_{N,i}+P_{F,j}\boldsymbol{\Phi}_{N,i})&\nonumber\\
&\times\boldsymbol{V}_{j})+\sigma^{2},&\label{pro29e}\\
&\bar{\Psi}_{F,i}\leq\sum\nolimits_{j=1,j\neq i}^{K}\mathrm{tr}((P_{N,j}\boldsymbol{\Phi}_{F,i}+P_{F,j}\boldsymbol{\Phi}_{F,i})&\nonumber\\
&\times\boldsymbol{V}_{j})+\sigma^{2},&\label{pro29f}
\end{align}\label{pro29}%
\end{subequations}
where $\boldsymbol{\Phi}_{\varepsilon,i}=\boldsymbol{W}^{\sss H}(\hat{\boldsymbol{h}}_{\varepsilon,i}+\boldsymbol{\Delta}\boldsymbol{h}_{\varepsilon,i})(\hat{\boldsymbol{h}}_{\varepsilon,i}+\boldsymbol{\Delta}\boldsymbol{h}_{\varepsilon,i})^{\sss H}\boldsymbol{W}$. Since the objective function in (\ref{pro29a}) is non-convex, we use the SCA method based on the first-order Taylor series to approximate (\ref{pro29a}), and problem (\ref{pro29}) is re-expressed as
\begin{subequations}
\begin{align}
\max_{\{\boldsymbol{V}_{i}\},\Psi_{\varepsilon,i},\bar{\Psi}_{\varepsilon,i}}&~\sum_{i=1}^{K}\omega\log_{2}\left(\bar{\Psi}_{N,i}+\Psi_{N,i}\right)-\omega\log_{2}((\bar{\Psi}_{N,i})^{(t-1)})\nonumber\\
&+\omega\frac{1}{\ln2 (\bar{\Psi}_{N,i})^{(t-1)}}(\bar{\Psi}_{N,i}-(\bar{\Psi}_{N,i})^{(t-1)})\nonumber\\
&+(1-\omega)\log_{2}\left(\bar{\Psi}_{F,i}+\Psi_{F,i}\right)-(1-\omega)\nonumber\\
&\log_{2}((\bar{\Psi}_{F,i})^{(t-1)})+(1-\omega)\frac{1}{\ln2 (\bar{\Psi}_{F,i})^{(t-1)}}\nonumber\\
&(\bar{\Psi}_{F,i}-(\bar{\Psi}_{F,i})^{(t-1)}),\label{pro30a}\\
\mbox{s.t.}~
&(\ref{pro29b}),(\ref{pro29c}), (\ref{pro29d}),(\ref{pro29e}),(\ref{pro29f}).&\label{pro30b}%
\end{align}\label{pro30}%
\end{subequations}
Problem (\ref{pro30}) can be solved using a semidefinite programming (SDP) algorithm. Specifically, a non-convex problem can be transformed into an SDP problem. The core idea is to solve the problem using existing methods, such as the interior point method. After solving the problem (\ref{pro30}) and obtaining its solution (expressed as eigenvalues), the optimal $\{\boldsymbol{V}_{i}\}$ (if $\{\boldsymbol{V}_{i}\}$ is rank one) can be derived through eigenvalue decomposition. Additionally, convex relaxation techniques, such as the SDP concave-convex procedure for phase beamforming, can be employed to further refine the solution quality. If the semidefinite relaxation (SDR) of the problem (\ref{pro30}) is not guaranteed to be tight, an approximate first-order solution can be obtained through iterative methods. Fortunately, it has been proven that the SDR of the problem (\ref{pro30}) is sometimes tight when the solution satisfies $\mathrm{Rank}(\boldsymbol{V}_{i})=1$. Thus, we can theoretically derive two specific conditions that validate $\mathrm{rank}(\boldsymbol{V}_{i}) = 1$  based on the proof by contradiction in\cite{10078317,10315058,10035459}.

\subsection{RHS beamforming optimization problem}
With the fixed $P_{\varepsilon,i}$, $\{\boldsymbol{v}_{i}\}$ and $\boldsymbol{\Delta}\boldsymbol{h}_{\varepsilon,i}$, and define $\mathrm{vec}((\hat{\boldsymbol{h}}_{\varepsilon,i}+\boldsymbol{\Delta}\boldsymbol{h}_{\varepsilon,i})^{\sss H}\boldsymbol{W}\boldsymbol{v}_{j})=(\boldsymbol{v}_{j}^{\sss T}\otimes (\hat{\boldsymbol{h}}_{\varepsilon,i}+\boldsymbol{\Delta}\boldsymbol{h}_{\varepsilon,i})^{\sss H})\mathrm{vec}(\boldsymbol{W})$, let $\boldsymbol{w}=\mathrm{vec}(\boldsymbol{W})$, $\mathrm{vec}((\hat{\boldsymbol{h}}_{\varepsilon,i}+\boldsymbol{\Delta}\boldsymbol{h}_{\varepsilon,i})^{\sss H}\boldsymbol{W}\boldsymbol{v}_{j})=(\boldsymbol{v}_{j}^{\sss T}\otimes (\hat{\boldsymbol{h}}_{\varepsilon,i}+\boldsymbol{\Delta}\boldsymbol{h}_{\varepsilon,i})^{\sss H})\boldsymbol{w}$, the RHS beamforming problem is given by
\begin{subequations}
\begin{align}
\max_{\boldsymbol{w}}&~\sum_{i=1}^{K}\omega\log_{2}\left(1+\frac{\Psi_{N,i}}{\bar{\Psi}_{N,i}}\right)+(1-\omega)\nonumber\\
&\times\log_{2}\left(1+\frac{\Psi_{F,i}}{\bar{\Psi}_{F,i}}\right),\label{pro31a}\\
\mbox{s.t.}~
&|(\boldsymbol{v}_{j}^{\sss T}\otimes (\hat{\boldsymbol{h}}_{\varepsilon,i}+\boldsymbol{\Delta}\boldsymbol{h}_{\varepsilon,i})^{\sss H})\boldsymbol{w}|^{2}\leq \epsilon_{\varepsilon,i,j},&\label{pro31b}\\
&\Psi_{\varepsilon,i}\geq P_{\varepsilon,i}|(\boldsymbol{v}_{i}^{\sss T}\otimes (\hat{\boldsymbol{h}}_{\varepsilon,i}+\boldsymbol{\Delta}\boldsymbol{h}_{\varepsilon,i})^{\sss H})\boldsymbol{w}|^{2},&\label{pro31d}\\
&\bar{\Psi}_{N,i}\leq\sum_{j=1,j\neq i}^{K}\left(P_{N,j}|((\boldsymbol{v}_{j}^{\sss T}\otimes (\hat{\boldsymbol{h}}_{N,i}+\boldsymbol{\Delta}\boldsymbol{h}_{N,i})^{\sss H})\boldsymbol{w}|^{2}\right.&\nonumber\\
&\left.+P_{F,j}|(\boldsymbol{v}_{j}^{\sss T}\otimes (\hat{\boldsymbol{h}}_{N,i}+\boldsymbol{\Delta}\boldsymbol{h}_{N,i})^{\sss H})\boldsymbol{w}|^{2}\right)+\sigma^{2},&\label{pro31f}\\
&\bar{\Psi}_{F,i}\leq\sum_{j=1,j\neq i}^{K}\left(P_{N,j}|(\boldsymbol{v}_{j}^{\sss T}\otimes (\hat{\boldsymbol{h}}_{F,i}+\boldsymbol{\Delta}\boldsymbol{h}_{F,i})^{\sss H})\boldsymbol{w}|^{2}\right.&\nonumber\\
&\left.+P_{F,j}|(\boldsymbol{v}_{j}^{\sss T}\otimes (\hat{\boldsymbol{h}}_{F,i}+\boldsymbol{\Delta}\boldsymbol{h}_{F,i})^{\sss H})\boldsymbol{w}|^{2}\right)+\sigma^{2},&\label{pro31g}
\end{align}\label{pro31}%
\end{subequations}    
We use the first-order Taylor series to obtain the upper bound $|(\boldsymbol{v}_{j}^{\sss T}\otimes (\hat{\boldsymbol{h}}_{\varepsilon,i}+\boldsymbol{\Delta}\boldsymbol{h}_{\varepsilon,i})^{\sss H})\boldsymbol{w}|^{2}$ and it is given in (\ref{pro32}) at the top of the this page.
\begin{figure*}
\begin{align}
|(\boldsymbol{v}_{t}^{\sss T}\otimes (\hat{\boldsymbol{h}}_{\varepsilon,i}+\boldsymbol{\Delta}\boldsymbol{h}_{\varepsilon,i})^{\sss H})\boldsymbol{w}|^{2}\geq& 2\mathrm{Re}\left\{(\boldsymbol{w}^{(n-1)})^{\sss H}(\boldsymbol{v}_{t}^{\sss T}\otimes (\hat{\boldsymbol{h}}_{\varepsilon,i}+\boldsymbol{\Delta}\boldsymbol{h}_{\varepsilon,i})^{\sss H})(\boldsymbol{v}_{t}^{\sss T}\otimes (\hat{\boldsymbol{h}}_{\varepsilon,i}+\boldsymbol{\Delta}\boldsymbol{h}_{\varepsilon,i})^{\sss H})^{\sss H}\boldsymbol{w}\right\}
\nonumber\\
&-|(\boldsymbol{v}_{t}^{\sss T}\otimes (\hat{\boldsymbol{h}}_{\varepsilon,i}+\boldsymbol{\Delta}\boldsymbol{h}_{\varepsilon,i})^{\sss H})^{\sss H}\times\boldsymbol{w}^{(n-1)}|^{2}=\Upsilon_{\varepsilon}(\boldsymbol{w}).\label{pro32}
\end{align}
\hrulefill 
\end{figure*}
Problem (\ref{pro31}) is rewritten as
\begin{subequations}
\begin{align}
\max_{\boldsymbol{w}}&~\sum_{i=1}^{K}\omega\log_{2}\left(1+\frac{\Psi_{N,i}}{\bar{\Psi}_{N,i}}\right)+(1-\omega)\nonumber\\
&\times\log_{2}\left(1+\frac{\Psi_{F,i}}{\bar{\Psi}_{F,i}}\right),\label{pro33a}\\
\mbox{s.t.}~
&(\ref{pro31b}),(\ref{pro31d}),&\label{pro33b}\\
&\bar{\Psi}_{N,i}\leq\sum\nolimits_{j=1,j\neq i}^{K}\left(P_{N,j}\Upsilon_{N}(\boldsymbol{w})+P_{F,j}\Upsilon_{N}(\boldsymbol{w})\right)\nonumber\\
&+\sigma^{2},&\label{pro33c}\\
&\bar{\Psi}_{F,i}\leq\sum\nolimits_{j=1,j\neq i}^{K}\left(P_{N,j}\Upsilon_{F}(\boldsymbol{w})+P_{F,j}\Upsilon_{F}(\boldsymbol{w})\right)\nonumber\\
&+\sigma^{2}.&\label{pro33d}
\end{align}\label{pro33}
\end{subequations}    
We use the SCA method to solve this problem (\ref{pro33}). 

\begin{algorithm}%
\caption{Beamforming Design Algorithm in the Inner Loop} \label{algo1}
\hspace*{0.02in}{\bf Initialize:} 
$\boldsymbol{W}^{(0)}$, $\boldsymbol{v}_{i}^{(0)}$, $(\boldsymbol{\Delta}\boldsymbol{h}_{\varepsilon,i})^{(0)}$. Determine $\epsilon_{\varepsilon,i}$, $\epsilon_{\boldsymbol{\Delta}\boldsymbol{h}_{\varepsilon,i}}$ based on (\ref{pro24}) and (\ref{pro27}).\\
\hspace*{0.02in}{\bf Repeat:}~$t=t+1$.\\
Given $\boldsymbol{W}^{(t)}$, $\boldsymbol{v}_{i}^{(t)}$, $P_{N,i}^{(t)}$,$P_{F,i}^{(t)}$, utilizing SCA and SDR for problem (\ref{pro28}). Update $(\boldsymbol{\Delta}\boldsymbol{h}_{\varepsilon,i})^{(t+1)}$;\\
Given $(\boldsymbol{\Delta}\boldsymbol{h}_{\varepsilon,i})^{(t+1)}$,  $\boldsymbol{W}^{(t)}$,  utilizing SCA for problem (\ref{pro30}). Update $\boldsymbol{v}^{(t+1)}$;\\
Given $(\boldsymbol{\Delta}\boldsymbol{h}_{\varepsilon,i})^{(t+1)}$, $\boldsymbol{v}^{(t+1)}$,  utilizing SCA for problem (\ref{pro33}). Update $\boldsymbol{W}^{(t+1)}$;\\
\hspace*{0.02in}{\bf Until:}~Convergence.\\
\hspace*{0.02in}{\bf Output:}
$(\boldsymbol{\Delta}\boldsymbol{h}_{\varepsilon,i})^{(t+1)}$, $\boldsymbol{v}^{(t+1)}$, $\boldsymbol{W}^{(t+1)}$.
\end{algorithm}

\section{Proposed algorithm for NU and FU power allocation}
In this section, we formulate the joint NU and FU power allocation (PA) problem in (\ref{pro13}). We find that the objective function on NU power $P_{N,i}$ and FU power $P_{F,i}$ is non-convex. To deal with this non-convex problem, we introduce the auxiliary variable method in\cite{9472958}. Specifically, let $P_{i}=P_{N,i}+P_{F,i}$ and $P_{i}$ denotes the group power, therefore, problem (\ref{pro13}) is re-expressed as
\begin{subequations}
\begin{align}
\max_{\{P_{i}\}}\max_{\{P_{N,i},P_{F,i}\}}&~\sum\nolimits_{i=1}^{K}\omega R_{N,i}+(1-\omega)\omega R_{F,i},\label{pro34a}\\
\mbox{s.t.}~
&(\ref{pro11g}),(\ref{pro11h}),(\ref{pro11k}),&\label{pro34b}\\
&P_{N,i}+P_{F,i}=P_{i},&\label{pro34c}\\
&\sum\nolimits_{i=1}^{K}P_{i}=P.&\label{pro34d}
\end{align}\label{pro34}%
\end{subequations}    
Since NF effective channel gain is larger than FF effective channel gain, i.e., $|(\hat{\boldsymbol{h}}_{N,i}+\boldsymbol{\Delta}\boldsymbol{h}_{N,i})^{\sss H}\boldsymbol{W}\boldsymbol{v}_{j}|^{2}\geq |(\hat{\boldsymbol{h}}_{F,i}+\boldsymbol{\Delta}\boldsymbol{h}_{F,i})^{\sss H}\boldsymbol{W}\boldsymbol{v}_{j}|^{2}$. Therefore, to maximize the transmission rate of the NUs and FUs, we set FU to meet the minimum rate requirement $R_{F,i}=\gamma_{F,i}$, and the remaining power is allocated to NU \cite{8790780}. Thus, we have
\begin{align}
&P_{N,i}=P_{i}-P_{F,i},\nonumber\\
&P_{F,i}=\frac{2^{\gamma_{F,i}}}{1+2^{\gamma_{F,i}}}\left(P_{i}+\frac{\sigma^{2}}{|(\hat{\boldsymbol{h}}_{F,i}+\boldsymbol{\Delta}\boldsymbol{h}_{F,i})^{H}\boldsymbol{W}\boldsymbol{v}_{j}|^{2}}\right).\label{pro35}
\end{align}    
Since $R_{F,i}=\gamma_{F,i}$
, we can rewrite (\ref{pro34a}) as
\begin{align}
&R=\sum\nolimits_{i=1}^{K}\omega R_{N,i}+(1-\omega)\sum\nolimits_{i=1}^{K}\gamma_{F,i}.\label{pro36}
\end{align}
As $(1-\omega)\sum\nolimits_{i=1}^{N}\gamma_{F,i}$ is constant, it can be rewritten as $\sum_{i=1}^{K}\omega R_{N,i}$. Thus, problem (\ref{pro34}) is rewritten as
\begin{subequations}
\begin{align}
\max_{\{P_{i}\}}&~\sum\nolimits_{i=1}^{\sss K}\omega R_{N,i},\label{pro37a}\\
\mbox{s.t.}~
&(\ref{pro34d}),&\label{pro37b}\\
&\log_{2}(1+\mathrm{SINR}_{N,i})\geq \gamma_{N,i}.&\label{pro37c}
\end{align}\label{pro37}%
\end{subequations}    
Since the objective function of $\sum_{i=1}^{K}\omega R_{i}^{\sss N}$ is non-convex, the PA problem of NUs still needs to be solved. To solve this problem,  we propose an iterative algorithm. Without (\ref{pro37c}), the optimization problem in (\ref{pro37}) is a convex problem. Therefore, we have the following theorem.
\begin{theorem}\label{the3}
Romving NU rate constraint (\ref{pro37c}), the globally optimal solution of problem (\ref{pro37}) is given by
\begin{align}
&\bar{P}_{i}=P/K-(\mu_{i}+1)/\kappa_{i}+\sum\nolimits_{j=1}^{K}(\alpha_{j}+1)/(K\beta_{j}),\label{pro38}
\end{align}
where
\begin{align}
&\kappa_{i}=|(\hat{\boldsymbol{h}}_{N,i}+\boldsymbol{\Delta}\boldsymbol{h}_{N,i})^{\sss H}\boldsymbol{W}\boldsymbol{v}_{j}|^{2}/\sigma^{2}\left(1/2^{\gamma_{F,i}}\right),\nonumber\\
&\mu_{i}=-|(\hat{\boldsymbol{h}}_{N,i}+\boldsymbol{\Delta}\boldsymbol{h}_{N,i})^{\sss H}\boldsymbol{W}\boldsymbol{v}_{j}|^{2}/|(\hat{\boldsymbol{h}}_{F,i}+\boldsymbol{\Delta}\boldsymbol{h}_{F,i})^{\sss H}\boldsymbol{W}\boldsymbol{v}_{j}|^{2}\nonumber\\
&\left(1/2^{\gamma_{F,i}}\right).\label{pro39}
\end{align}
The proof is given in \textbf{Appendix \ref{appB}}.
\end{theorem} 
Based on \textbf{Theorem~\ref{the3}}, if $\bar{P}_{i}$ in (\ref{pro38})  meets the constraints in (\ref{pro37c}), then $\bar{P}_{1,i}$ is the optimal solution of (\ref{pro37}). However, if $\bar{P}_{i}$ fails to satisfy the constraint in (\ref{pro37c}), $\bar{P}_{i}$ is not the optimal solution of (\ref{pro37}). In this latter scenario, the result in \textbf{Theorem~\ref{the4}} can be applied.

\begin{theorem}\label{the4}
If (\ref{pro37c}) is taken into account, the globally optimal solution should consistently meet the constraint in (\ref{pro40}) at the top of this page.
\begin{figure*}[ht]
\begin{align}
\hat{P}_{1,i}&=\frac{2^{\gamma_{N,i}}-\mu_{i}}{\kappa_{i}},\nonumber \\\forall i&\in\left\{i|1\leq i\leq K, \frac{|(\hat{\boldsymbol{h}}_{N,i}+\boldsymbol{\Delta}\boldsymbol{h}_{N,i})^{H}\boldsymbol{W}\boldsymbol{v}_{j}|^{2}P_{i,2}}{|(\hat{\boldsymbol{h}}_{N,i}+\boldsymbol{\Delta}\boldsymbol{h}_{N,i})^{H}\boldsymbol{W}\boldsymbol{v}_{j}|^{2}P_{i,1}+\sigma^{2}}<2^{\gamma_{N,i}}-1, \frac{|(\hat{\boldsymbol{h}}_{F,i}+\boldsymbol{\Delta}\boldsymbol{h}_{F,i})^{H}\boldsymbol{W}\boldsymbol{v}_{j}|^{2}P_{i,1}}{|(\hat{\boldsymbol{h}}_{F,i}+\boldsymbol{\Delta}\boldsymbol{h}_{F,i})^{H}\boldsymbol{W}\boldsymbol{v}_{j}|^{2}P_{i,2}+\sigma^{2}}<2^{\gamma_{F,i}}-1\right\}.\label{pro40}%
\end{align}
\hrulefill 
\end{figure*}
The proof is given in \textbf{Appendix \ref{appC}}. 
\end{theorem}
When $i\in\left\{i|1\leq i\leq K, \frac{|(\hat{\boldsymbol{h}}_{N,i}+\boldsymbol{\Delta}\boldsymbol{h}_{N,i})^{H}\boldsymbol{W}\boldsymbol{v}_{j}|^{2}P_{F,i}}{|(\hat{\boldsymbol{h}}_{N,i}+\boldsymbol{\Delta}\boldsymbol{h}_{N,i})^{H}\boldsymbol{W}\boldsymbol{v}_{j}|^{2}P_{N,i}+\sigma^{2}}\right.$
$\left.\leq2^{\gamma_{F,i}}-1\right\}$, the optimal PA can be obtained by solving the following problem
\begin{subequations}
\begin{align}
\max_{\{P_{i}\}}&~\sum_{i=1}^{K}\omega R_{N,i},\label{pro41a}\\
\mbox{s.t.}~
&R_{N,i}\geq \gamma_{N,i},&\label{pro41b}\\
&\sum_{i\notin\mathcal{K}}P_{i}\leq P-\sum_{j\notin\mathcal{K}}\hat{P}_{j}.&\label{pro41c}
\end{align}\label{pro41}%
\end{subequations}    
Based on (\ref{pro41}), the proposed algorithms follow these steps. Initially, we address the issue in (\ref{pro41}) while disregarding the constraint (\ref{pro41b}), allowing the use of \textbf{Theorem~\ref{the3}}. Next, \textbf{Theorem~\ref{the4}} is applied to find solutions that do not comply with constraint (\ref{pro41b}), followed by updating problem (\ref{pro41}). This process is repeated until it converges.
The PA algorithm is summarized in Algorithm 2.
\begin{algorithm}%
\caption{PA Algorithm in the Outer loop} \label{algo3-1}
\hspace*{0.02in}{\bf Initialization:}
$t=0$, $P_{i}^{(t)}=P_{i}^{(0)}=\frac{P}{2}$.\\
\hspace*{0.02in}{\bf Repeat:}\\
$\mathcal{V}=\{1,2,\cdots,K\}$, let $\mathcal{K}=\mathcal{V}$.\\
\hspace*{0.02in}{\bf If:}~$\mathcal{K}\neq\emptyset$\\
\hspace*{0.02in}{\bf Repeat:}\\
Based on \textbf{Theorem~\ref{the3}}, calculate $\kappa_{i}$, $\mu_{i}$, and $\bar{P}_{i}^{(t)}$.\\
Update $\mathcal{K}$. Based on \textbf{Theorem~\ref{the4}}, calculate $\hat{P}_{i}^{(t)}$.\\
Update the set based on $\mathcal{V}=\mathcal{V}/\mathcal{K}$.\\
\hspace*{0.02in}{\bf Until:}
$\mathcal{K}=\emptyset$.\\
\hspace*{0.02in}{\bf Update:} $\hat{P}_{i}^{(t)}=\bar{P}_{i}^{(t)}$.\\
\hspace*{0.02in}{\bf Until:}
$P_{i}^{*}=\hat{P}_{i}^{(T_{max})}$, where $T_{max}$ is the maximum number of iterations.\\
\hspace*{0.02in}{\bf Output:}
$\{P_{i}^{*}\}$.\\
\end{algorithm}

\begin{figure}[htbp]
  \centering
  \includegraphics[scale=0.5]{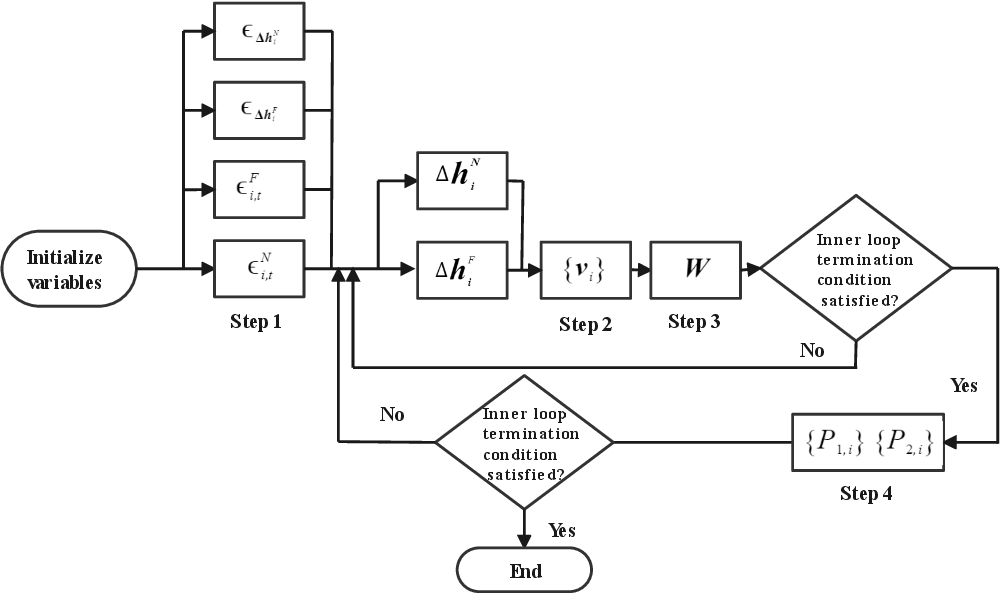}
  \caption{ The flow chart exposing the internal structure of the proposed algorithm. }\vspace{-10pt}
\label{FIGURE1}
\end{figure}

\subsection{Computational Complexity Analysis}
According to Fig.\ref{FIGURE1}, the proposed algorithm consists of four steps. 
In \textbf{step~1}, the optimization problem (\ref{pro23}) is solved by using the SCA method. Since there are $2K$ constraints in the problem (\ref{pro24}), the number of iterations that are required for the SCA method is
$\mathcal{O}(\sqrt{2K}\log_{2}(\frac{1}{\epsilon_{1}}))$\cite{10068140}, where $\epsilon_{1}$ is the accuracy of the SCA method for solving problem
(\ref{pro24}). Similarly, in \textbf{step~2}, the number of iterations that are required for the SCA algorithm for the problem (\ref{pro27}) is $\mathcal{O}(\sqrt{8K}\log_{2}(\frac{1}{\epsilon_{2}}))$. In \textbf{step~3}, the number of iterations that are required for the SCA method is
$\mathcal{O}(\sqrt{7K}\log_{2}(\frac{1}{\epsilon_{3}}))$, where $\epsilon_{3}$ is the accuracy of the SCA method for solving problem
(\ref{pro24}). The number of out-loop iteration is $\kappa_{in}(\mathcal{O}(\sqrt{2K}\log_{2}(\frac{1}{\epsilon_{1}}))+\mathcal{O}(\sqrt{8K}\log_{2}(\frac{1}{\epsilon_{2}}))+\mathcal{O}(\sqrt{7K}\log_{2}(\frac{1}{\epsilon_{1}})))$, where $\kappa_{in}$ is the number of inner-loop iterations. In \textbf{step~4}, the complexity of calculating the effective channel gains of the users is
$\mathcal{KN_{T}M}$. Each time that $\{P_{N,i}\}$ and $\{P_{F,i}\}$ are updated, the maximum number of iterations is $M$, and the
complexity of computing ${P_{N,i},P_{F,i}}$ in each subcycle is no larger than $\mathcal{O}(K^{2})$. Thus, the complexity
of PA algorithm is $\mathcal{O}(KN_{T}M+\kappa_{out}MK^{2})$, where $\kappa_{out}$ is the number of outer-loop iterations. The total computing complexity of the proposed algorithm is $\mathcal{O}(\kappa_{in}(\sqrt{2K}\log_{2}(\frac{1}{\epsilon_{1}})+\sqrt{8K}\log_{2}(\frac{1}{\epsilon_{2}})+\sqrt{7K}\log_{2}(\frac{1}{\epsilon_{3}}))+KN_{T}M+\kappa_{out}MK^{2})$. In this simulation, we set $\epsilon_{1}$, $\epsilon_{2}$ and $\epsilon_{3}$ as $0.001$ according to\cite{8294044}.

\section{Numerical Results}\label{V}
This section provides numerical simulation results to validate the effectiveness of the proposed joint beamforming design and PA strategies. 
We assume the central element of the DMA is positioned at the origin of the coordinate system. The NUs and FUs are randomly placed on circular rings with radii of $10$ meters (m) and $15$~$m$, respectively. The number of DMA reflecting elements is $M=32$. The total transmit power is $P=27~dBm$, and the noise power at the receiver is $\sigma^{2}=-80~dBm$. The carrier frequency is $f=60~GHz$, and the speed of light is $3\times10^{8}~m/s$. The signal wavelength is determined by $\bar{\lambda}=\frac{c}{f}$. For large-scale path loss, we use the values of $\zeta_{0}=-30~dB$, $d_{0}=1~m$, and $\alpha=2.2$\cite{9324910,zhang2024near}. The Rician factor is  $\kappa_{R}=20$. The $l_{2}$-norm of $\boldsymbol{h}_{i}^{NLoS}$ and $\boldsymbol{h}_{i}^{LoS}$ are set to $\delta^{NLoS}=10^{-4}$ and $\delta^{LoS}=0$, respectively. Additionally, the user position error bound and the number of reflecting elements $N$ (or $L$) will be varied for different observations. Based on\cite{10070578}, we set $\epsilon_{\boldsymbol{\Delta}\boldsymbol{p}_{N,i}}$ and $\epsilon_{\boldsymbol{\Delta}\boldsymbol{p}_{F,i}}$ as $0.1$~m.


\begin{figure}[htbp]
\centering
\begin{minipage}[t]{0.48\textwidth}
\centering
\includegraphics[width=0.8\textwidth,height=0.6\textwidth]{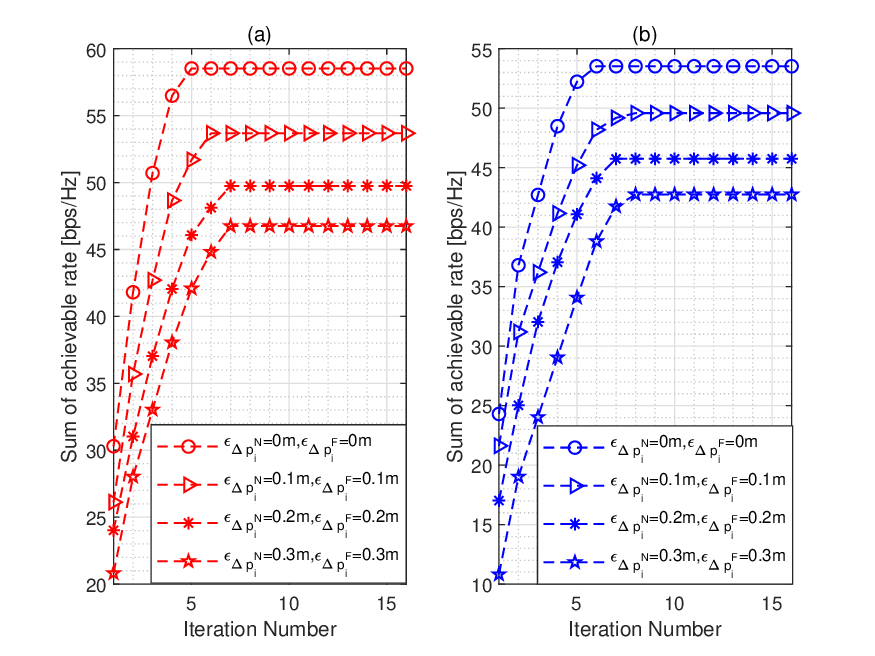}
\caption{Transmit power versus the number of iterations.}
\label{FIGURE2}
\end{minipage}
\end{figure}
Fig.~\ref{FIGURE2} shows the convergence performance of the two-layer algorithm and the BCD algorithm as a function of the number of iterations. As the number of iterations increases, the system and rate values show a monotonically increasing trend and converge to a stable solution in about $4$ to $8$ iterations. This further proves the convergence of the proposed two-layer algorithm and BCD algorithm. In addition, when the positioning error increases, the system sum rate can still converge after $4$ to $8$ iterations, which shows that the proposed algorithm has better robust performance.

\begin{figure}[htbp]
\centering
\begin{minipage}[t]{0.48\textwidth}
\centering
\includegraphics[scale=0.45]{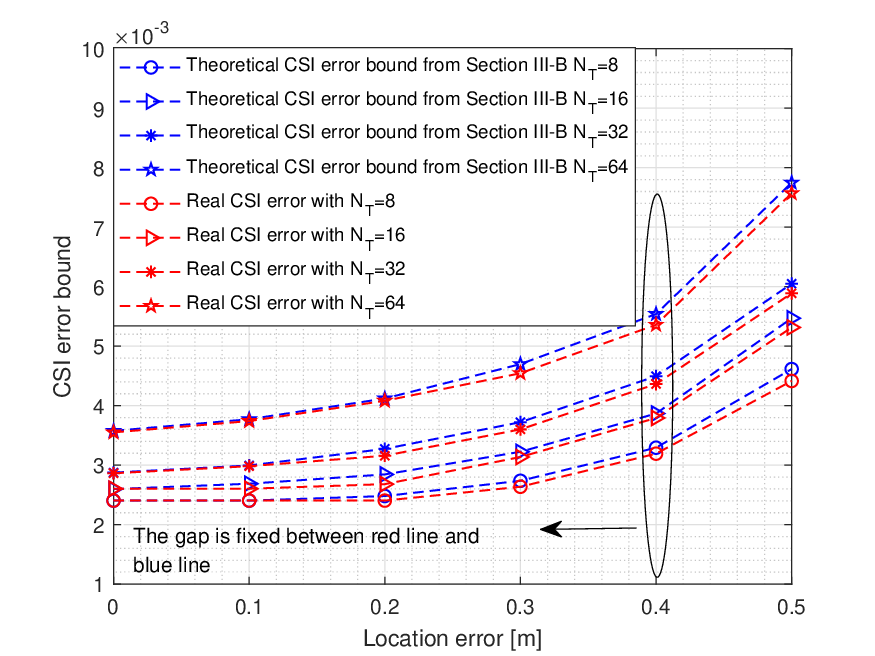}
\caption{Transmit power versus IoT device harvested power threshold.}
\label{FIGURE4}
\end{minipage}
\end{figure}

\begin{figure}[htbp]
\begin{minipage}[t]{0.48\textwidth}
\centering
\includegraphics[scale=0.45]{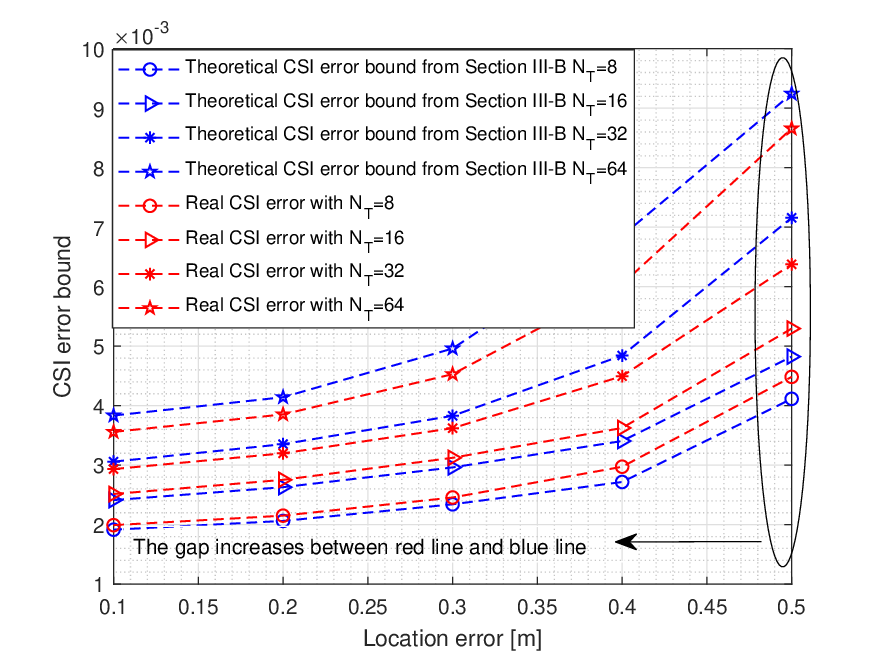}
\caption{Transmit power versus IoT device SINR threshold.}
\label{FIGURE3}
\end{minipage}
\end{figure}
Fig.~\ref{FIGURE4} depicts the CSI error bound of NU with respect to the number of DMA reflective elements and $\Delta\boldsymbol{p}_{i}^{\sss N}$. This shows that, first of all, the overall theoretical results of NU agree well with the simulation results, which means that the derivation of the CSI error bound is accurate. Finally, the CSI error bound increases with larger $N_{T}$ or $\Delta\boldsymbol{p}_{i}^{\sss N}$, which illustrates that more reflective elements or a larger degree of NU position uncertainty will lead to more severe CSI errors. Fig.~\ref{FIGURE3} depicts the CSI error bound range of the FU with respect to the number of DMA reflective elements and $\Delta\boldsymbol{p}_{i}^{\sss F}$, the overall theoretical results of FU are in poor agreement with the simulation results after the position error of $\Delta\boldsymbol{p}_{i}^{\sss F}>0.2m$. This is because the channel of the NU only considers the LoS channel, while the channel of the FU considers the influence of NLoS, which means that the CSI error bound of FU The derivation is correct. This shows that the CSI error bound of the FU increases with larger $N_{T}$ or $\Delta\boldsymbol{p}_{i}^{\sss F}$, which illustrates that more reflective elements or a larger degree of FU position uncertainty will lead to a more severe CSI error of FU.

\begin{figure}[htbp]
\centering
\begin{minipage}[t]{0.48\textwidth}
\centering
\includegraphics[scale=0.45]{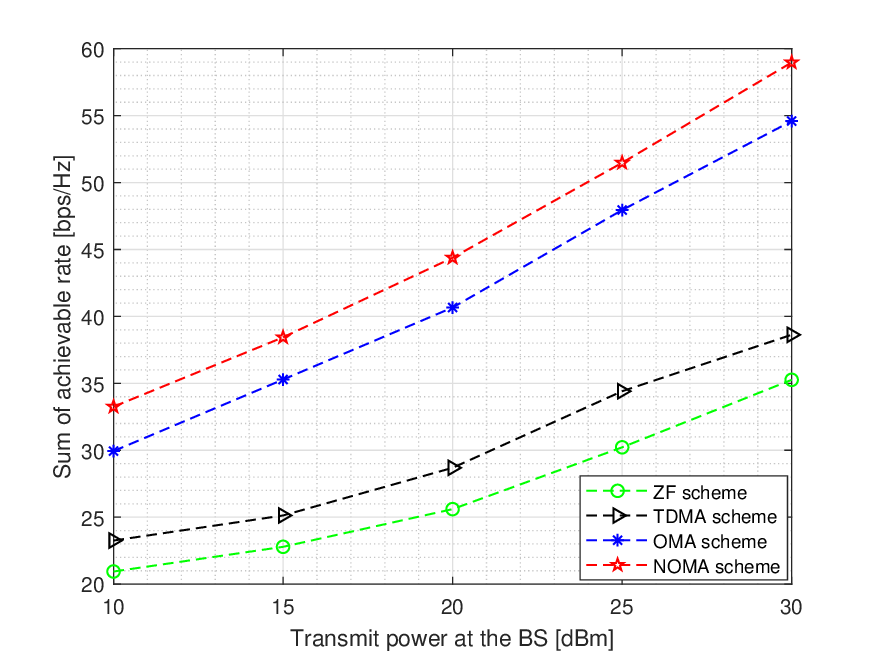}
\caption{Transmit power versus the number of iterations.}
\label{FIGURE5}
\end{minipage}
\begin{minipage}[t]{0.48\textwidth}
\centering
\includegraphics[scale=0.45]{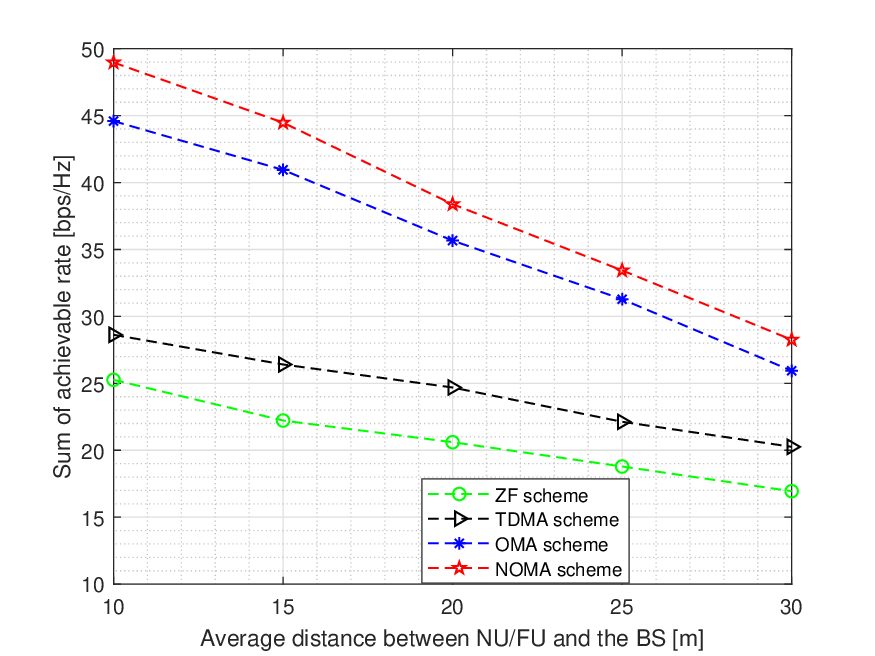}
\caption{Transmit power versus IoT device SINR threshold.}
\label{FIGURE6}
\end{minipage}
\end{figure}
In Fig.\ref{FIGURE5}, we compare the communication performance of the proposed scheme with four position error baselines. We examine the variations in transmit power and the average distance from NU and FU to the BS separately, assuming a fixed distance of $5$~$m$ between NU and FU. Additionally, simultaneous changes in their positions are considered to adjust the average distance to the BS. The results indicate that as the positioning error increases, the system's total sum rate decreases. This can be explained by the fact that the power and beam allocation of the NU and FU are unreasonable due to increased positioning error, and mismatch between NU and FU based channels and actual channels, which deteriorates the communication performance of the network. In addition, it can be found that as the transmission distance increases, the total reachable rate shows a downward trend. This is because the greater the transmission distance, the greater the path loss, thereby obtaining more transmit power to maintain the same rate. Otherwise, the rate will decrease.

In Fig.~\ref{FIGURE6}, we compare the communication performance of the proposed approach with four baseline methods. We examine variations in transmit power and the average distance from an NU and an FU to the BS. The results reveal that, among all the approaches, the NOMA scheme utilizing beam steering achieves the highest sum achievable rate. The following facts can explain this: 1) NOMA allows the BS to serve the NU and FU in the same time-frequency resource block through flexible power control, which is capacity-enabled and thus able to achieve better performance the OMA scheme (i.e., frequency division address and time division multiple access); 2) The proposed algorithm is robust to position errors and can achieve higher system performance; 3) Since the traditional ZF scheme is designed only based on NU's CSI, inevitably the channel leading to the FU is very weak. Therefore, the FU should be allocated more power to meet its quality of service requirements, while NU will receive less power, which limits the communication rate of the network.

\begin{figure}[htbp]
\begin{minipage}[t]{0.48\textwidth}
\centering
\includegraphics[scale=0.45]{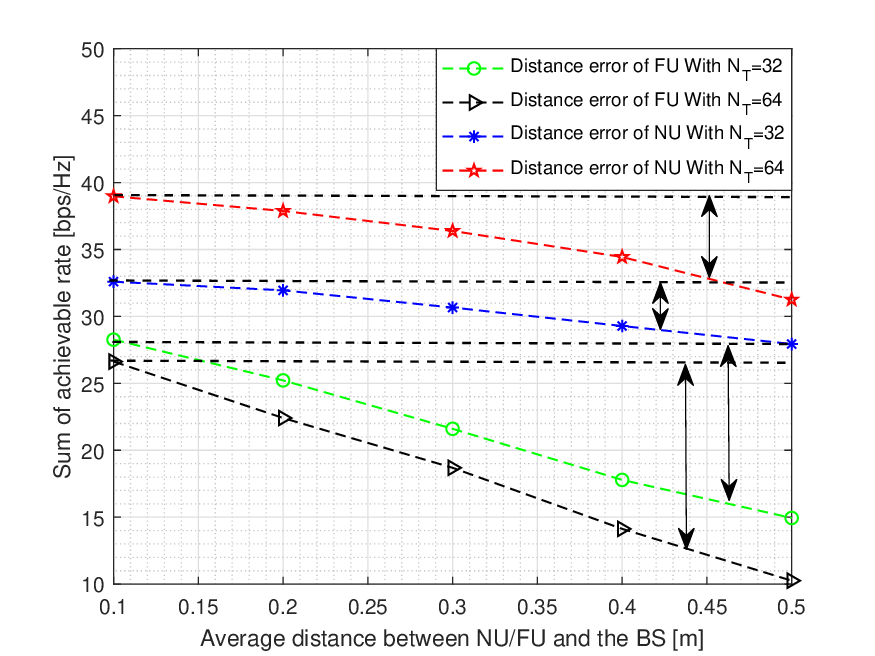}
\caption{Transmit power versus IoT device SINR threshold.}
\label{FIGURE7}
\end{minipage}
\end{figure}
As shown in Fig.\ref{FIGURE7}, the total reachability rate of the NOMA network decreases with the increase in average position error. This is expected, as imperfect distance knowledge leads to a discrepancy between the beam pattern and the actual spherical wave channel, resulting in degraded network performance. However, an interesting result emerges: the proposed robust beamforming design scheme is not sensitive to distance estimation errors of NU and FU. This is because we account for positioning errors under both PA and beam design strategies, enhancing system performance and rate while meeting all worst-case constraints. However, when the distance error exceeds 0.16m, the performance of our algorithm also declines. As the position error increases, the NU and FU channels based on position parameters cannot match the actual channels, leading to a failure to meet the power requirements for their respective QoS. Consequently, the decrease in NU channel gain due to imperfect distance information significantly affects the network's achievable rate.

\section{Conclusion and Future Works}\label{VI}
We propose an NF NOMA transmission framework that supports DMA. NOMA was used to enhance the transmission connectivity of the overloaded network, and the user position was used to obtain CSI. The approximate CSI error bound in user position uncertainty. Then, to mitigate the impact of position errors on beamforming design, we formulated a worst-case robust beamforming optimization problem. To effectively address this non-convex issue, we proposed an iterative optimization method utilizing Lagrange multipliers and matrix inverse lemmas, evaluating its performance in terms of worst-case SNR for the UE, convergence behavior, and algorithm efficiency. The results demonstrate that the proposed beamforming scheme outperforms other baseline approaches, and the near-field NOMA transmission framework supporting DMA is more resilient to NU's CSI distance estimation errors. Compared to existing robust beamforming techniques based on the OMA, TDMA, and ZF schemes, our algorithm achieves better performance and faster convergence to the optimal value.

\appendices
\section{Proof of the \textbf{Theorem~\ref{th1}}}\label{appA}
According to (\ref{pro23}), the biggest challenge of the derivation of (\ref{pro23}) in the problem are $\cos\left(\frac{2\pi}{\lambda}\Theta_{\varepsilon,v+(h-1)N_{T,h}}\right)$. To deal with this challenge, we use the Taylor series to  approximate $\cos\left(\frac{2\pi}{\lambda}\Theta_{\varepsilon,v+(h-1)N_{T,h}}\right)$, we have
\begin{align}
&\cos\left(\frac{2\pi}{\lambda}\Theta_{\varepsilon,v+(h-1)N_{T,h}}\right)\approx 1-\frac{2\pi^{2}}{\lambda^{2}}(\Theta_{\varepsilon,v+(h-1)N_{T,h}})^{2}\nonumber\\
&+\frac{2\pi^{4}}{3\lambda^{4}}(\Theta_{\varepsilon,v+(h-1)N_{T,h}})^{4}.\label{proA1}
\end{align}
$\Xi(\Delta\boldsymbol{p}_{\varepsilon,i})$ can be rewritten as (\ref{pro_A5}) on the top of this page.
\begin{figure*}[ht] 
\centering 
\vspace*{8pt} 
\begin{align}
\Xi(\Delta\boldsymbol{p}_{\varepsilon,i})\approx&\sum\nolimits_{v=1}^{N_{T,v}}\sum\nolimits_{h=1}^{N_{T,h}}
\left\{\left\|\boldsymbol{x}_{v+(h-1)N_{T,h}}-\hat{\boldsymbol{p}}_{\varepsilon,i}-\Delta\boldsymbol{p}_{N,i}\right\|_{2}^{-\alpha}+\left\|\boldsymbol{x}_{v+(h-1)N_{T,h}}-\hat{\boldsymbol{p}}_{\varepsilon,i}\right\|_{2}^{-\alpha}\right\}-2\sum\nolimits_{v=1}^{N_{T,v}}\sum\nolimits_{h=1}^{N_{T,h}}\nonumber\\
&\left\{\left\|\boldsymbol{x}_{v+(h-1)N_{T,h}}\right.\right.\left.\left.-\hat{\boldsymbol{p}}_{\varepsilon,i}-\Delta\boldsymbol{p}_{\varepsilon,i}\right\|_{2}^{-\frac{\alpha}{2}}\left\|\boldsymbol{x}_{v+(h-1)N_{T,h}}-\hat{\boldsymbol{p}}_{\varepsilon,i}\right\|_{2}^{-\frac{\alpha}{2}}\left(1-(2\pi^{2}/\lambda^{2})(\Theta_{\varepsilon,v+(h-1)N_{T,h}})^{2}+\right.\right.\nonumber\\
&\left.\left.(2\pi^{4}/3\lambda^{4})(\Theta_{\varepsilon,v+(h-1)N_{T,h}})^{4}\right)\right\}.\label{pro_A5}
\end{align}
\hrulefill 
\end{figure*}%
Then, we have 
\begin{align}
\cos\left(\frac{2\pi}{\lambda}\Theta_{\varepsilon, v+(h-1)N_{T,h}}\right)\leq 1.\label{proA2}
\end{align}
According to (\ref{proA2}), the following inequality is given by
\begin{align}
&\sum_{\varepsilon\in\mathcal{C}}\frac{4\pi^{2}}{\lambda^{2}}(\Theta_{\varepsilon,v+(h-1)N_{T,h}})^{2}-\frac{4\pi^{4}}{3\lambda^{4}}(\Theta_{\varepsilon,v+(h-1)N_{T,h}})^{4}\nonumber\\
&\geq 0.\label{proA3}%
\end{align}
Based on the triangle inequality, we have
\begin{align}
&\left\|\boldsymbol{x}_{v+(h-1)N_{T,h}}-\hat{\boldsymbol{p}}_{\varepsilon,i}-\Delta\boldsymbol{p}_{\varepsilon,i}\right\|_{2}^{-\frac{\alpha}{2}}\leq \left(\left\|\boldsymbol{x}_{v+(h-1)N_{T,h}}\right.\right.\nonumber\\
&\left.\left.-\hat{\boldsymbol{p}}_{\varepsilon,i}\right\|_{2}\right.\left.-\epsilon_{\Delta\boldsymbol{p}_{i, \varepsilon}}\right)^{-\frac{\alpha}{2}}.\label{proA4}%
\end{align}
Since $\alpha$ is positive, the upper bound of $\sum_{\varepsilon\in\mathcal{C}}\Xi(\boldsymbol{\Delta}\boldsymbol{p}_{\varepsilon,i})$ are further given in (\ref{proA5}) on the top of the next page.
\begin{figure*}[ht] 
\centering 
\vspace*{8pt} 
\begin{align}
&\sum_{\varepsilon\in\mathcal{C}}\Xi(\boldsymbol{\Delta}\boldsymbol{p}_{\varepsilon,i})\leq \sum_{\varepsilon\in\mathcal{C}}\Psi_{1}(\boldsymbol{\Delta}\boldsymbol{p}_{\varepsilon,i})+\sum_{\varepsilon\in\mathcal{C}}\Sigma_{2}(\boldsymbol{\Delta}\boldsymbol{p}_{\varepsilon,i})+\sum_{\varepsilon\in\mathcal{C}}\Sigma_{\varepsilon,i}((2\pi^{2}/\lambda^{2})(\Theta_{\varepsilon,v+(h-1)N_{T,h}})^{2}-(2\pi^{4}/3\lambda^{4})(\Theta_{\varepsilon,v+(h-1)N_{T,h}})^{4}).\label{proA5}
\end{align}
\hrulefill 
\end{figure*}%
where $\Psi(\boldsymbol{\Delta}\boldsymbol{p}_{\varepsilon,i})=\sum_{v=1}^{N_{T,v}}\sum_{h=1}^{N_{T,h}}
\left\{\left\|\boldsymbol{x}_{v+(h-1)N_{T,h}}-\hat{\boldsymbol{p}}_{\varepsilon,i}-\boldsymbol{\Delta}\boldsymbol{p}_{i, \varepsilon}\right\|_{2}^{-\alpha}+\right.$\\
$\left.\left\|\boldsymbol{x}_{v+(h-1)N_{T,h}}-\hat{\boldsymbol{p}}_{\varepsilon,i}\right\|_{2}^{-\alpha}\right\}$,
$\Sigma(\boldsymbol{\Delta}\boldsymbol{p}_{i, \varepsilon})=2\sum_{v=1}^{N_{T,v}}\sum_{h=1}^{N_{T,h}}\left\{\left\|\boldsymbol{x}_{v+(h-1)N_{T,h}}-\hat{\boldsymbol{p}}_{\varepsilon,i}-\boldsymbol{\Delta}\boldsymbol{p}_{i, \varepsilon}\right\|_{2}^{-\frac{\alpha}{2}}\right.$\\
$\left.\right.\left.\left\|\boldsymbol{x}_{v+(h-1)N_{T,h}}-\hat{\boldsymbol{p}}_{\varepsilon,i}\right\|_{2}^{-\frac{\alpha}{2}}\right\}$.  $\Sigma_{\varepsilon,i}$ is denoted as
\begin{align}
&\Sigma_{\varepsilon,i}=\left(\left\|\boldsymbol{x}_{v+(h-1)N_{T,h}}-\hat{\boldsymbol{p}}_{\varepsilon,i}\right\|_{2}-\epsilon_{\boldsymbol{\Delta}\boldsymbol{p}_{i, \varepsilon}}\right)^{-\frac{\alpha}{2}}\nonumber\\
&\left\|\boldsymbol{x}_{v+(h-1)N_{T,h}}-\hat{\boldsymbol{p}}_{\varepsilon,i}\right\|_{2}^{-\frac{\alpha}{2}}.\label{proA_5}%
\end{align}
It is not difficult to find $\sum_{\varepsilon\in\mathcal{C}}\Psi(\boldsymbol{\Delta}\boldsymbol{p}_{\varepsilon,i})+\sum_{\varepsilon\in\mathcal{C}}\Sigma_{1}(\boldsymbol{\Delta}\boldsymbol{p}_{\varepsilon,i})$ is non-linear and non-convex over $\boldsymbol{\Delta}\boldsymbol{p}_{\varepsilon,i}$. Then, by using second-order Taylor approximation to deal with $\sum_{\varepsilon\in\mathcal{C}}\Psi(\boldsymbol{\Delta}\boldsymbol{p}_{\varepsilon,i})+\sum_{\varepsilon\in\mathcal{C}}\Sigma_{1}(\boldsymbol{\Delta}\boldsymbol{p}_{\varepsilon,i})$ and have
\begin{align}
&\sum_{\varepsilon\in\mathcal{C}}\Psi(\boldsymbol{\Delta}\boldsymbol{p}_{\varepsilon,i})+\sum_{\varepsilon\in\mathcal{C}}\Sigma_{1}(\boldsymbol{\Delta}\boldsymbol{p}_{\varepsilon,i})\approx \Psi_{1}(\boldsymbol{0})+\Psi_{2}(\boldsymbol{0})+\Sigma_{1}(\boldsymbol{0})\nonumber\\
&+\Sigma_{2}(\boldsymbol{0})+\frac{\partial \Psi_{1}(\boldsymbol{\Delta}\boldsymbol{p}_{N,i})+\Sigma_{1}(\boldsymbol{\Delta}\boldsymbol{p}_{N,i})}{\partial \boldsymbol{\Delta}\boldsymbol{p}_{N,i}}|_{\boldsymbol{\Delta}\boldsymbol{p}_{N,i}=0}\boldsymbol{\Delta}\boldsymbol{p}_{N,i}\nonumber\\
&+\frac{1}{2}(\boldsymbol{\Delta}\boldsymbol{p}_{N,i})^{T}\frac{\partial}{\partial (\boldsymbol{\Delta}\boldsymbol{p}_{N,i})^{T}}\left(\frac{\partial\Psi_{1}(\boldsymbol{\Delta}\boldsymbol{p}_{N,i})+\Sigma_{1}(\boldsymbol{\Delta}\boldsymbol{p}_{N,i})}{\partial \boldsymbol{\Delta}\boldsymbol{p}_{N,i}}\right.\nonumber\\
&\left.|_{\boldsymbol{\Delta}\boldsymbol{p}_{N,i}=0}\right)\boldsymbol{\Delta}\boldsymbol{p}_{N,i}.\label{proA6}%
\end{align}
Next, we continue to deal with $\frac{2\pi^{4}}{3\lambda^{4}}(\Theta_{\varepsilon,v+(h-1)N_{T,h}})^{4}$. Based on the power mean inequality, we have
\begin{align}
&-\Sigma_{\varepsilon,i}\frac{4\pi^{4}}{3\lambda^{4}N}(\Theta_{\varepsilon,v+(h-1)N_{T,h}})^{4}\leq\nonumber\\
&-\frac{4\pi^{4}}{3\lambda^{4}N}((\boldsymbol{\Delta}\boldsymbol{p}_{\varepsilon,i})^{T}\boldsymbol{\Omega}_{\varepsilon,i}\boldsymbol{\Delta}\boldsymbol{p}_{\varepsilon,i})^{2}.\label{proA7}%
\end{align}%
The upper bound of $\sum_{\varepsilon\in\mathcal{C}}\Xi(\boldsymbol{\Delta}\boldsymbol{p}_{\varepsilon,i})$ is given by 
\begin{align}
\sum_{\varepsilon\in\mathcal{C}}(\boldsymbol{\Delta}\boldsymbol{p}_{\varepsilon,i})^{T}\boldsymbol{\Upsilon}_{\varepsilon,i}\boldsymbol{\Delta}\boldsymbol{p}_{\varepsilon,i}-\frac{4\pi^{4}}{3\lambda^{4}N}((\boldsymbol{\Delta}\boldsymbol{p}_{\varepsilon,i})^{T}\boldsymbol{\Omega}_{\varepsilon,i}\boldsymbol{\Delta}\boldsymbol{p}_{\varepsilon,i})^{2}.\label{proA8}%
\end{align}
Then, we introduce auxiliary variables $\varpi_{\varepsilon,i}$ and $\hat{\varpi}_{\varepsilon,i}$, the optimization problem is given by
\begin{subequations}
\begin{align}
\max_{\boldsymbol{\Delta}\boldsymbol{p}_{\varepsilon,i},\varpi_{\varepsilon,i},\hat{\varpi}_{\varepsilon,i}}&~-\sum_{i=1}^{K}(4\pi^{4}/3\lambda^{4}N)((\varpi_{N,i})^{2}+(\varpi_{F,i})^{2})\nonumber\\
&+\hat{\varpi}_{N,i}+\hat{\varpi}_{F,i},\label{proA9a}\\
\mbox{s.t.}~
&(\ref{pro23b}),&\\
&\varpi_{\varepsilon,i}=(\boldsymbol{\Delta}\boldsymbol{p}_{\varepsilon,i})^{H}\boldsymbol{\Omega}_{\varepsilon,i}\boldsymbol{\Delta}\boldsymbol{p}_{\varepsilon,i},
&\label{proA9b}\\
&\hat{\varpi}_{\varepsilon,i}=(\boldsymbol{\Delta}\boldsymbol{p}_{\varepsilon,i})^{H}\boldsymbol{\Upsilon}_{\varepsilon,i}\boldsymbol{\Delta}\boldsymbol{p}_{\varepsilon,i},
&\label{proA9c}
\end{align}\label{proA9}%
\end{subequations}%
Then, non-convex constraints (\ref{proA9b}) and (\ref{proA9c}) are equivalently denoted as $\varpi_{\varepsilon,i}\leq(\boldsymbol{\Delta}\boldsymbol{p}_{\varepsilon,i})^{H}\boldsymbol{\Omega}_{\varepsilon,i}\boldsymbol{\Delta}\boldsymbol{p}_{\varepsilon,i}$, $\varpi_{\varepsilon,i}\geq(\boldsymbol{\Delta}\boldsymbol{p}_{\varepsilon,i})^{H}\boldsymbol{\Omega}_{\varepsilon,i}\boldsymbol{\Delta}\boldsymbol{p}_{\varepsilon,i}$, $\hat{\varpi}_{\varepsilon,i}\leq(\boldsymbol{\Delta}\boldsymbol{p}_{\varepsilon,i})^{H}\boldsymbol{\Upsilon}_{\varepsilon,i}\boldsymbol{\Delta}\boldsymbol{p}_{\varepsilon,i}$ and $\hat{\varpi}_{\varepsilon,i}\geq(\boldsymbol{\Delta}\boldsymbol{p}_{\varepsilon,i})^{H}\boldsymbol{\Upsilon}_{\varepsilon,i}\boldsymbol{\Delta}\boldsymbol{p}_{\varepsilon,i}$. Thus, problem (\ref{proA9}) is rewritten as
\begin{subequations}
\begin{align}
\max_{\boldsymbol{\Delta}\boldsymbol{p}_{\varepsilon,i},\varpi_{\varepsilon,i},\hat{\varpi}_{\varepsilon,i}}&~-\sum_{i=1}^{K}(4\pi^{4}/3\lambda^{4}N)((\varpi_{N,i})^{2}+(\varpi_{F,i})^{2})\nonumber\\
&+\hat{\varpi}_{N,i}+\hat{\varpi}_{F,i},\label{proA10a}\\
\mbox{s.t.}~
&(\ref{pro23b}),&\\
&\varpi_{\varepsilon,i}\geq(\boldsymbol{\Delta}\boldsymbol{p}_{\varepsilon,i})^{H}\boldsymbol{\Omega}_{\varepsilon,i}\boldsymbol{\Delta}\boldsymbol{p}_{\varepsilon,i},
&\label{proA10b}\\
&\hat{\varpi}_{\varepsilon,i}\geq(\boldsymbol{\Delta}\boldsymbol{p}_{\varepsilon,i})^{H}\boldsymbol{\Upsilon}_{\varepsilon,i}\boldsymbol{\Delta}\boldsymbol{p}_{\varepsilon,i},
&\label{proA10c}\\
&\varpi_{\varepsilon,i}\leq(\boldsymbol{\Delta}\boldsymbol{p}_{\varepsilon,i})^{H}\boldsymbol{\Omega}_{\varepsilon,i}\boldsymbol{\Delta}\boldsymbol{p}_{\varepsilon,i},&\label{proA10d}\\
&\hat{\varpi}_{\varepsilon,i}\geq(\boldsymbol{\Delta}\boldsymbol{p}_{\varepsilon,i})^{H}\boldsymbol{\Upsilon}_{\varepsilon,i}\boldsymbol{\Delta}\boldsymbol{p}_{\varepsilon,i},&\label{proA10e}
\end{align}\label{proA10}%
\end{subequations}%
Finally, we use the SCA method to deal with the non-convex constraints (\ref{proA10d}) and (\ref{proA10e}). Therefore, problem (\ref{proA10}) is written as in (\ref{pro24}).
The proof is completed.

\section{Proof of \textbf{Theorem~\ref{the3}}}\label{appB}
Because the inter-group interference is very small, the inter-group interference can be ignored\cite{9472958}. Therefore, the SINR for NU in $i$-th group is re-expressed as follows:
\begin{align}
u_{N,i}=\frac{|(\hat{\boldsymbol{h}}_{N,i}+\boldsymbol{\Delta}\boldsymbol{h}_{N,i})^{H}\boldsymbol{W}\boldsymbol{v}|^{2}}{\sigma^{2}}.\label{proC1}
\end{align}
Based on $P_{N,i}+P_{F,i}=P_{i}$, the $i$-th group total power $P_{i}$ is a linear function with respect to NF power $P_{N,i}$. Therefore, $u_{i}$ can be rewritten as a function of $P_{i}$, and $u_{i}$ is given by
\begin{align}
u_{N,i}=\kappa_{i}P_{i}+\mu_{i},\label{proC2}
\end{align}
where $\kappa_{i}$ and $\mu_{i}$ are expressed as (\ref{pro39}).
We find that $\kappa_{i}\geq 0$ and $\mu_{i}\geq 0$. Then, (\ref{pro34d}) and (\ref{pro37c}) are reformulated as
\begin{align}
\sum\nolimits_{i=1}^{K}\log_{2}(\kappa_{i}P_{i}+\mu_{i}+1), \lambda_{i,1}-\mu_{i}-P_{i}\kappa_{i}\leq 0.\label{proC4}
\end{align}
According to (\ref{proC4}), the problem in (\ref{pro37}) can be denoted as
\begin{subequations}
\begin{align}
\max_{\{P_{i}\}}&~\sum_{i\in\mathcal{K}}\omega R_{N,i},\label{proC5a}\\
\mbox{s.t.}~
&\lambda_{i,1}-\mu_{i}-P_{i}\kappa_{i}\leq 0,~ \sum\nolimits_{i=1}^{K}P_{i}\leq 0&\label{proC5b}
\end{align}\label{proC5}%
\end{subequations}    
We denote the objective function in (\ref{proC5a}) as $g(\{P_{i}\})=\sum_{i=1}^{K}\log_{2}(\kappa_{i}P_{i}+\mu_{i})$. We find that $g(\{P_{i}\})$ is a monotonically increasing function of $P_{n}$. By ignoring the constraint (\ref{pro37c}), problem (\ref{pro37}) can be solved by the Karush-Kuhn-Tucker (KKT) conditions, i.e.,
\begin{align}
\frac{\partial g(\{P_{i}\})}{\partial P_{i}}=\varrho, \sum\nolimits_{i=1}^{K}P_{i}=P.\label{proC6}
\end{align}
The solution of the equation in (\ref{proC6}) is given in (\ref{pro38}).
Then, \textbf{Theorem~\ref{the3}} is proved.

\section{Proof of \textbf{Theorem~\ref{the4}}}\label{appC}
To prove \textbf{Theorem~\ref{the4}}, we use the method of the proof by contradiction. $\hat{P}_{N_{1},i}$ is set as the optimal solution, and $\hat{P}_{N_{1},i}$ satisfies $\hat{P}_{N_{1}}>\frac{2^{\gamma_{N,i}}-\mu_{i}}{\kappa_{i}}>\bar{P}_{N_{1}}$. Since $\sum_{k=1}^{K}\hat{P}_{k}\leq P$, there always exists $\hat{P}_{N_{2}}\leq \bar{P}_{N_{2}}(N_{2}\neq N_{1})$. The PA solution can be expressed as
\begin{align}
T_{N_{1}}=\hat{P}_{N_{1}}-\delta, T_{N_{2}}=\hat{P}_{N_{2}}+\delta, T_{i}=\hat{P}_{i}, i\neq N_{1},N_{2}.\label{proD1}
\end{align}
Therefore, we only need to prove that the sum rate of the PA solution $\{T_{i}\}$ is
higher than that of the solution $\{\hat{P}_{i}\}$. We assume that the objective function in (\ref{pro37a}) with
solution $\{T_{i}\}$ is $f(\{T_{i}\}) = \sum_{i=1}^{K}\log_{2}
(\kappa_{i}T_{i}+\mu_{i})$, and the objective function in (\ref{pro37a}) with
solution $\{\hat{P}_{i}\}$ is $q(\{\hat{P}_{i}\}) =\sum_{i=1}^{K}\log_{2}(\kappa_{i}\hat{P}_{i}+\mu_{i})$. If $\varpi=0$, then $f(\{T_{i}\})-q(\{\hat{P}_{i}\})=0$. The
derivative of $f(\{T_{i}\})-q(\{\hat{P}_{i}\})$ with respect to $\varpi$ is
\begin{align}
&\frac{\partial f(\{T_{i}\})}{\partial \varpi}-\frac{\partial q(\{\hat{P}_{i}\})}{\partial \varpi}=\frac{1}{\ln 2}\frac{\kappa_{N_{2}}}{(\kappa_{N_{2}}(\hat{P}_{N_{2}}+\varpi)+\mu_{N_{2}}+1)}\nonumber\\
&-\frac{1}{\ln 2}\frac{\kappa_{N_{1}}}{(\kappa_{N_{1}}(\hat{P}_{N_{1}}-\varpi)+\mu_{N_{1}}+1)}.\label{proD2}
\end{align}
According to the derivative of the objective function in (\ref{pro37a}), we have
\begin{align}
&\frac{1}{\ln 2}\frac{\kappa_{N_{2}}}{(\kappa_{N_{2}}(\hat{P}_{N_{2}}+\varpi)+\mu_{N_{2}}+1)}>\frac{1}{\ln 2}\frac{\kappa_{N_{2}}}{(\kappa_{N_{2}}\bar{P}_{N_{2}}+\mu_{N_{2}}+1)}\nonumber\\
&\frac{1}{\ln 1}\frac{\kappa_{N_{1}}}{(\kappa_{N_{1}}(\hat{P}_{N_{1}}+\varpi)+\mu_{N_{1}}+1)}<\frac{1}{\ln 2}\frac{\kappa_{N_{1}}}{(\kappa_{N_{1}}\bar{P}_{N_{1}}+\mu_{N_{1}}+1)},\label{proD3}
\end{align}
where $\frac{1}{\ln 2}\frac{\kappa_{N_{1}}}{(\kappa_{N_{1}}\bar{P}_{N_{1}}+\mu_{N_{1}}+1)}=\frac{1}{\ln 2}\frac{\kappa_{N_{2}}}{(\kappa_{N_{2}}\bar{P}_{N_{2}}+\mu_{N_{2}}+1)}=\frac{1}{\ln 2}\frac{N}{P+\sum_{i=1}^{K}\frac{\mu_{i}+1}{\kappa_{i}}}$. Therefore, (\ref{proD3}) is equivalent to
\begin{align}
&\frac{\partial f(\{T_{i}\})}{\partial \varpi}-\frac{\partial q(\{\hat{P}_{i}\})}{\partial \varpi}>\frac{1}{\ln 2}\frac{\kappa_{N_{2}}}{(\kappa_{N_{2}}\bar{P}_{N_{2}}+\mu_{N_{2}}+1)}-\nonumber\\
&\frac{1}{\ln 2}\frac{\kappa_{N_{1}}}{(\kappa_{N_{1}}\bar{P}_{N_{1}}+\mu_{N_{1}}+1)}=0.\label{proD4}
\end{align}
Since $f(\{T_{i}\})-q(\{\hat{P}_{i}\})=0$, we have $f(\{T_{i}\})>q(\{\hat{P}_{i}\})$, which demonstrates that $\{T_{i}\}$ is better than $\{\hat{P}_{i}\}$. This contradicts the assumption that $\{\hat{P}_{i}\}$ is the optimal solution. To this end, the optimal solution of problem (\ref{pro37}) must satisfy $\hat{P}_{i}=\frac{2^{\gamma_{N,i}}-\mu_{i}}{\kappa_{i}}$. Thus, \textbf{Theorem~\ref{the4}} is
proved.



\end{document}